%% file: Bs2K.tex
\documentclass[preprint,aps,prd,floatfix,preprintnumbers,eqsecnum,nofootinbib,superscriptaddress]{revtex4-1}
\usepackage[T1]{fontenc}
\usepackage[utf8]{inputenc}
\usepackage{dcolumn}
\usepackage{amsmath,amssymb}
\usepackage{color}
\usepackage{bm}
\usepackage{enumitem}
\usepackage{rotating}
\usepackage[colorlinks=true,linkcolor=blue,filecolor=blue,urlcolor=blue,citecolor=blue,pdftex,plainpages=false]{hyperref}
\input{macro}

\begin{document}
\input{title}
\input{section1-intro}
\input{section2-form-factors}
\input{section3-lqcd}

\input{section4-analysis}

\input{section5-errors}

\input{section6-continuum-form-factors}

\input{section7-pheno}

\input{section8-summary}
\input{acknowledgements}
\input{appendix}

\clearpage
\bibliographystyle{apsrev4-1}
\bibliography{Bs2K}
\nocite{*}

\end{document}

%% file: macro.tex
\graphicspath{{./figures/}}

\def\beq{\begin{equation}}
\def\enq{\end{equation}}
\def\fpar{f_\parallel}
\def\fperp{f_\perp}
\def\MBs{M_{B_s}}
\def\bpi{B \to \pi\ell\nu}
\def\bsk{B_s \to K\ell\nu}
\def\bd{B \to D \ell\nu}
\def\bsds{B_s \to D_s \ell\nu}
\def\tcut{t_\mathrm{cut}}

\renewcommand\Re{\operatorname{Re}}

%% file: title.tex
\preprint{FERMILAB-PUB-19-005-T}

\title{\boldmath $B_s\to K\ell\nu$ decay from lattice QCD}

\author{A.~Bazavov}
\affiliation{Department of Computational Mathematics, Science and Engineering, \\
and Department of Physics and Astronomy, Michigan State University, East Lansing, Michigan 48824, USA}

\author{C.~Bernard}
\affiliation{Department of Physics, Washington University, St.~Louis, Missouri 63130, USA}

\author{C.~DeTar}
\affiliation{Department of Physics and Astronomy, University of Utah, \\ Salt Lake City, Utah 84112, USA}

\author{Daping Du}
\affiliation{Department of Physics, Syracuse University, Syracuse, New York 13244, USA}

\author{A.X.~El-Khadra}
\affiliation{Department of Physics, University of Illinois, Urbana, Illinois 61801, USA} 
\affiliation{Fermi National Accelerator Laboratory, Batavia, Illinois 60510 USA}

\author{E.D.~Freeland}
\affiliation{Liberal Arts Department, School of the Art Institute of Chicago, Chicago, Illinois, USA}

\author{E.~G\'amiz}
\affiliation{CAFPE and Departamento de Fisica Te\'orica y del Cosmos, Universidad de Granada, E-18071 Granada, Spain}

\author{Z.~Gelzer}
\affiliation{Department of Physics, University of Illinois, Urbana, Illinois 61801, USA} 

\author{Steven~Gottlieb}
\affiliation{Department of Physics, Indiana University, Bloomington, Indiana 47405 USA}

\author{U.M.~Heller}
\affiliation{American Physical Society, Ridge, New York 11961, USA}

\author{A.S.~Kronfeld}
\affiliation{Fermi National Accelerator Laboratory, Batavia, Illinois 60510 USA}
\affiliation{Institute for Advanced Study, Technische Universit\"at M\"unchen, 85748 Garching, Germany}

\author{J.~Laiho}
\affiliation{Department of Physics, Syracuse University, Syracuse, New York 13244, USA}

\author{Yuzhi Liu}
\affiliation{Department of Physics, Indiana University, Bloomington, Indiana 47405 USA}

\author{P.B.~Mackenzie}
\affiliation{Fermi National Accelerator Laboratory, Batavia, Illinois 60510 USA}

\author{Y. Meurice}
\affiliation{Department of Physics and Astronomy, University of Iowa, \\ Iowa City, IA, USA}

\author{E.T.~Neil}
\affiliation{Department of Physics, University of Colorado, Boulder, Colorado 80309, USA}
\affiliation{RIKEN-BNL Research Center, Brookhaven National Laboratory, \\ Upton, New York 11973, USA}

\author{J.N.~Simone}
\affiliation{Fermi National Accelerator Laboratory, Batavia, Illinois 60510 USA}

\author{D.~Toussaint}
\affiliation{Physics Department, University of Arizona, Tucson, Arizona 85721, USA}

\author{R.S.~Van~de~Water}
\affiliation{Fermi National Accelerator Laboratory, Batavia, Illinois 60510 USA}

\author{Ran Zhou}
\affiliation{Fermi National Accelerator Laboratory, Batavia, Illinois 60510 USA}

\collaboration{Fermilab Lattice and MILC Collaborations}
\noaffiliation

\date{\today}

\include{abstract}

\maketitle
\let\endtitlepage\relax
\clearpage
\tableofcontents

%% file: abstract.tex
\begin{abstract}
We use lattice QCD to calculate the form factors $f_+(q^2)$ and $f_0(q^2)$ for
the semileptonic decay $B_s\to K\ell\nu$. Our calculation uses six MILC asqtad
2+1 flavor gauge-field ensembles with three lattice spacings.  At the smallest
and largest lattice spacing the light-quark sea mass is set to 1/10 the
strange-quark mass. At the intermediate lattice spacing, we use four values for
the light-quark sea mass ranging from 1/5 to 1/20 of the strange-quark mass. We
use the asqtad improved staggered action for the light valence quarks, and the
clover action with the Fermilab interpolation for the heavy valence bottom
quark. We use SU(2) hard-kaon heavy-meson rooted staggered chiral perturbation
theory to take the chiral-continuum limit. A functional $z$~expansion is used
to extend the form factors to the full kinematic range. We present predictions
for the differential decay rate for both $B_s\to K\mu\nu$ and $B_s\to
K\tau\nu$. We also present results for the forward-backward asymmetry, the
lepton polarization asymmetry, ratios of the scalar and vector form factors for
the decays  $B_s\to K\ell\nu$ and  $B_s\to D_s \ell\nu$. Our results, together
with future experimental measurements, can be used to determine the magnitude
of the Cabibbo-Kobayashi-Maskawa matrix element~$|V_{ub}|$.
\end{abstract}

%% file: section1-intro.tex
\section{Introduction} \label{sec:intro}

Semileptonic decays of hadrons can be used to determine elements of the
Cabibbo-Kobayashi-Maskawa (CKM) matrix. However, since the quarks that
participate in the underlying electroweak transition are constituents of bound
states, it is necessary to understand the effects of the strong interactions on
the decay. These effects are encapsulated in form factors for hadronic matrix
elements of the weak currents that govern the decay. Lattice QCD has allowed us
to calculate the form factors with increasing precision, making possible
stringent tests of the Standard Model and the CKM paradigm. Should there be a
violation of unitarity of the CKM matrix, or should two decay processes that
depend on the same CKM matrix element imply different values for that CKM
matrix element, we would have evidence for physics beyond the Standard Model.
The decay studied here, $\bsk$ depends on the same matrix element $V_{ub}$ as
the decay $\bpi$. Indeed, the only difference between the two decay processes
is that the light spectator up ($u$) or down ($d$) quark in the latter process
is replaced by a strange ($s$) quark in the case at hand. Since in lattice QCD,
strange quarks generally yield smaller statistical errors and are easier to
deal with computationally, a lattice calculation of the form factors for $\bsk$
decay can enable a precise $|V_{ub}|$ determination. This, in turn, can provide
a useful test of $|V_{ub}|$ determinations from the exclusive $\bpi$ and
$\Lambda_b \to p \ell \nu$ processes, and, if consistent, a reduced error on
$|V_{ub}|$ (exclusive) after combination. 
 
On the experimental side, however, while
BaBar~\cite{delAmoSanchez:2010af,Lees:2012vv} and
Belle~\cite{Ha:2010rf,Sibidanov:2013rkk} have published precise measurements of
the differential decay rate for $\bpi$, no such measurements exist
yet for $\bsk$. The branching fraction of the former decay is $(7.80 \pm
0.27)\times10^{-5}$~\cite{Tanabashi:2018oca}, as it is Cabibbo suppressed
compared to final states with charm. As BaBar and Belle observed many more $B
\bar B$ than $B_s \bar B_s$ events, it is not surprising that experimental
measurements of the latter decay have not yet been reported. In contrast, the
LHCb experiment at the CERN LHC collider observes decays of all $b$-flavored
hadrons, including $B_s$ mesons. They are expected to publish the results of
their ongoing $\bsk$ decay study within the coming
year~\cite{Ciezarek:2016lqu}. The Belle II experiment~\cite{Urquijo:2015qsa},
where the $e^+e^-$ collisions provide a cleaner environment than at the LHC,
also expects to study this decay. The current plans are that Belle II will
collect about $50~\text{ab}^{-1}$ at the $\Upsilon(4S)$ resonance (which decays
predominantly into $B$-meson pairs), and $5~\text{ab}^{-1}$ at the
$\Upsilon(5S)$, a rich source of $B_s$-meson pairs \cite{Urquijo:2015qsa}.
Thus, we do not expect the experimental accuracy for Belle~II's future
measurement of $\bsk$ decay rates to rival that of their expected results for
$\bpi$, but we do expect this decay to be studied by Belle~II.

This work is part of a broad study of flavor physics by the Fermilab Lattice
and MILC Collaborations to determine a number of CKM matrix elements from
semileptonic $K$~\cite{Bazavov:2012cd}, $D_{(s)}$~\cite{Aubin:2004ej}, and
$B_{(s)}$~\cite{Bernard:2008dn,%
Bailey:2008wp,Bailey:2012jg,Bailey:2012rr,Bailey:2014tva,Lattice:2015rga,Lattice:2015tia,Bailey:2015nbd,Bailey:2015dka,Du:2015tda}
decays using the asqtad $2+1$ flavor ensembles generated by the MILC
Collaboration~\cite{Bernard:2001av,Aubin:2004wf,Bazavov:2009bb}. These studies
are currently being
extended~\cite{Bazavov:2013maa,Gamiz:2016bpm,Primer:2017xzs,Gelzer:2017edb,Bazavov:2018kjg}
to use HISQ $2+1+1$ flavor ensembles~\cite{Bazavov:2010ru,Bazavov:2012xda}.
These newer ensembles include ones with physical-mass Goldstone pions at
several lattice spacings that significantly improve our control of the chiral
limit. In order to provide a systematic mode by mode comparison of results
obtained with the two sets of configurations, it is important to complete this
analysis of $\bsk$.

The techniques used here are very similar to those employed in
Ref.~\cite{Lattice:2015tia}, where the functional $z$ expansion was introduced.
However, in this work, we use a subset of six MILC ensembles covering a range
of lattice spacing $a$ between approximately 0.12 and 0.06 fm. Prior work used
12 ensembles including one with $a\approx 0.045$~fm.

The decay $\bsk$ has been studied by three other lattice-QCD groups, the HPQCD
Collaboration~\cite{Bouchard:2014ypa}, the RBC and UKQCD
Collaborations~\cite{Flynn:2015mha}, and the ALPHA
Collaboration~\cite{Bahr:2016ayy}, each choosing different actions for the
$b$-quark and for the light sea and valence quarks. Other previous calculations
of the $\bsk$ decay form factors are based on the relativistic quark
model~\cite{Faustov:2013ima}, light-cone sum
rules~\cite{Duplancic:2008tk,Khodjamirian:2017fxg}, and next-to-leading-order
(NLO) perturbative QCD~\cite{Wang:2012ab}. In
Sec.~\ref{subsec:compare_form_factors}, we compare our results with the prior
results. Preliminary reports on this study can be found in
Refs.~\cite{Liu:2013sya} and~\cite{Lattice:2017vqf}, where the vector current
renormalization factors were still multiplied by a blinding factor. This factor
was disclosed only after the analysis was finalized. 

The rest of this paper is organized as follows. In Sec.~\ref{sec:matrix}, we
define the continuum decay form factors and the hadronic matrix elements needed
to calculate them. In Sec.~\ref{sec:lattice}, we introduce the lattice QCD
operators and the form factors most convenient to calculate on the lattice. We
detail how to calculate the needed lattice matrix elements and enumerate the
MILC asqtad 2+1 flavor ensembles we have used. Section~\ref{sec:analysis}
discusses our analysis of the two- and three-point functions needed to
construct the lattice form factors. We also explain how we take the
chiral-continuum limit. Section~\ref{sec:errors} contains our analysis of
systematic errors in the range of momentum transfer accessible in our
calculation. To construct the continuum form factors over the entire range of
momentum transfer, we present the functional $z$ expansion in
Sec.~\ref{sec:contin}. We then apply it to obtain our final results for the
form factors. Section~\ref{sec:pheno} presents some of the phenomenological
implications of the results. Appendix~\ref{app:su2-chiral} contains details of
our application of SU(2) chiral perturbation theory to perform the chiral
extrapolation in Sec.~\ref{sec:analysis}. Appendix~\ref{app:reconstruct}
details how we construct the continuum form factors in
Sec.~\ref{subsec:continuum_form_factors}. Appendix~\ref{app:bin} contains the
binned differential decay rates, as well as the full correlation matrices.

%% file: section2-form-factors.tex
\section{Matrix elements and form factors} \label{sec:matrix}

\begin{figure}[b]
\centering
\includegraphics[width=0.47\columnwidth]{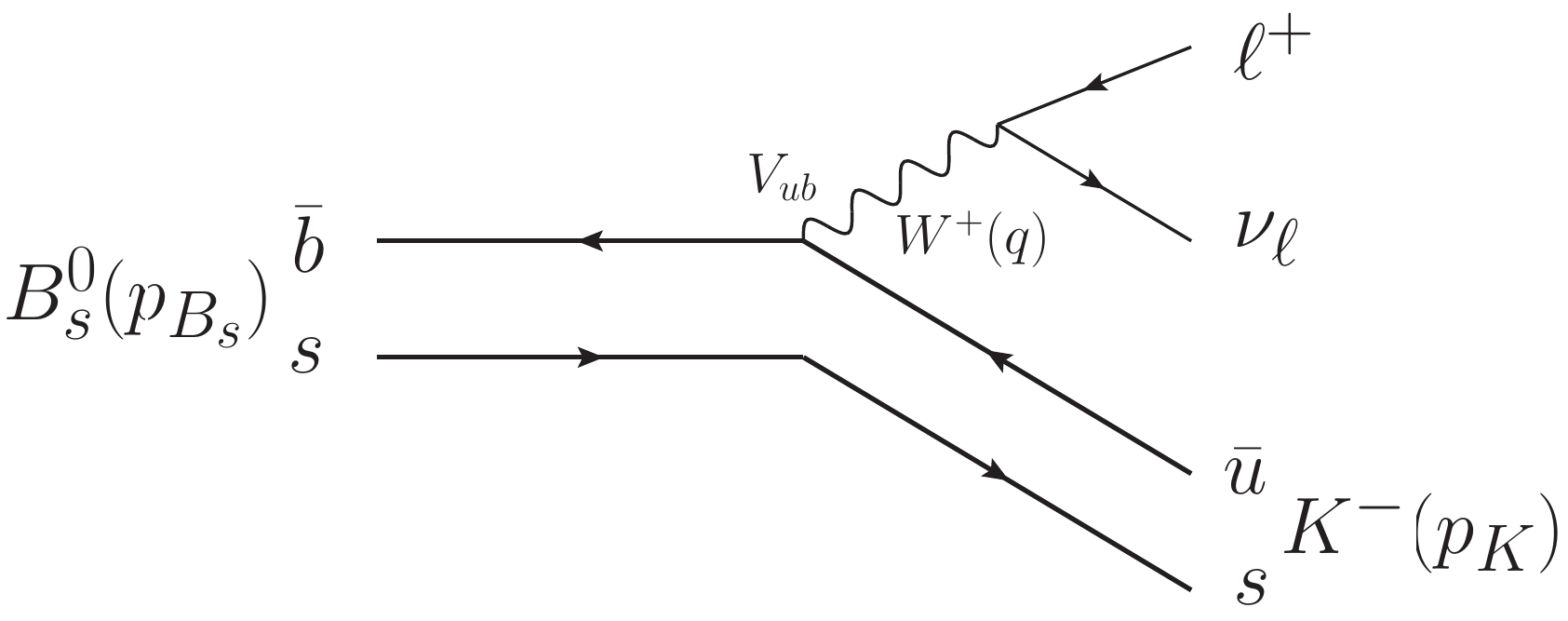}
\caption{Lowest order Standard Model Feynman diagram shown here for example of
semileptonic $B_s^0 \to K^- \ell^+\nu_\ell$ decay.}
\label{fig:diagram_with_momentum}
\end{figure}

To lowest order in the weak coupling constant, the semileptonic $\bsk$ decay
can be described via the Feynman diagram shown in
Fig.~\ref{fig:diagram_with_momentum}. 
The relevant hadronic matrix element can be written as
\beq
\left<K(p_K)|\mathcal{V}^\mu|B_s(p_{B_s})\right> =
\left(p^\mu_K + p^\mu_{B_s} -q^\mu\frac{\MBs^2 - M_K^2}{q^2}\right)
f_+(q^2) + q^\mu\frac{\MBs^2 - M_K^2}{q^2}f_0(q^2),
\label{eq:matrix_element_cont}
\enq
where $\mathcal{V}^\mu \equiv \bar{u}\gamma^\mu b$ is the vector current,
$p^\mu_{B_s}$ and $p^\mu_K$ are the $B_s$ and $K$ four-momenta, respectively,
$\MBs$ and $M_K$ are the corresponding meson masses, $q^\mu= p_{B_s}^\mu -
p_K^\mu$ is the momentum transferred to the lepton pair, and $f_+(q^2)$ and
$f_0(q^2)$ are the vector and scalar form factors corresponding to the exchange
of $1^-$ and $0^+$ particles. These two form factors are subject to a kinematic
constraint:
\beq
f_+(0) = f_0(0) ,
\label{eq:kin_constraint}
\enq
which eliminates the spurious pole at $q^2 = 0$ in
Eq.~(\ref{eq:matrix_element_cont}). The tensor form factor $f_T$ parametrizes
the hadronic matrix element of the tensor current
$T^{\mu\nu}=i\bar{u}\sigma^{\mu\nu}b$. Since it does not contribute to the
Standard Model decay rate, we do not include it in this calculation.

In the Standard Model, the angular-dependent differential decay rate for the
$\bsk$ can be written as
\begin{align}
\frac{d^2\Gamma}{dq^2 d \cos\theta_\ell}
=&
\frac{G_F^2 |V_{ub}|^2}{128\pi^3 M_{B_s}^2}
\left( 1 - \frac{m_\ell^2}{q^2} \right)^2
|\mathrm{p}_K|
\left[
4M_{B_s}^2 |\mathrm{p}_K|^2
\left(
\sin^2\theta_{\ell} + \frac{m_{\ell}^2}{q^2} \cos^2\theta_\ell
\right) 
|f_+(q^2)|^2 \right.
\nonumber \\
&+
\frac{4m_\ell^2}{q^2} (M_{B_s}^2 - M_K^2) M_{B_s} |\mathrm{p}_K|
\cos\theta_\ell 
\Re\left[f_+(q^2)  f_0^*(q^2)\right]
\nonumber \\
&\left. +
\frac{m_\ell^2}{q^2} (M_{B_s}^2 - M_K^2)^2 |f_0(q^2)|^2
\right] 
\label{eq:ddr}
\end{align}
in the $B_s$ meson rest frame. Here $G_F$ is the Fermi constant, $V_{ub}$ is an
element of the CKM matrix, $m_\ell$ is the lepton mass, and $\theta_\ell$ is
the angle between the final charged-lepton and the $B_s$ meson momenta in the
rest frame of the final state leptons. Thus, to determine $|V_{ub}|$ from a
measurement of the differential decay rate, it is necessary to compute the form
factor $f_+(q^2)$. If the charged lepton is the $\tau$, however, the lepton
mass cannot be neglected and $f_0(q^2)$ is also necessary. 

%% file: section3-lqcd.tex
\section{Lattice-QCD calculation} \label{sec:lattice}

In this section, we present the ingredients of our lattice-QCD calculation. The
definitions of form factors and correlation functions are given in
Sec.~\ref{subsec:def-formfactors}. The lattice actions and simulation
parameters are described in Sec.~\ref{subsec:action}. The lattice interpolating
operators, currents, and correlation functions are presented in
Sec.~\ref{subsec:operators}. 

\input{./sec3/subsection-3.1-defs}

\input{./sec3/subsection-3.2-actions}

\input{./sec3/subsection-3.3-currents}

%% file: sec3/subsection-3.1-defs.tex
\subsection{Definitions} \label{subsec:def-formfactors}

For lattice calculations and Heavy Quark Effective Theory (HQET), it is
convenient to work in the $B_s$ rest frame and introduce the $B_s$
four-velocity
\beq
v^\mu = p^\mu_{B_s}/\MBs .
\label{eq:p_M}
\enq
The square of the lepton momentum transfer $q^2$ can then be expressed as
\begin{align}
q^2 &=
(p_{B_s}^\mu - p_K^\mu)^2 = M_{B_s}^2 + M_K^2 - 2 \MBs E_K ,
\label{eq:qsq_E}
\end{align}
where $E_K = p_K \cdot v$ is the kaon energy. Defining
\beq
p^\mu_{\perp} \equiv p^\mu_K - (p_K \cdot v)v^\mu
\enq
as the projection of the kaon momentum in the direction perpendicular to
$v^\mu$ and using Eq.~(\ref{eq:qsq_E}), one can rewrite the matrix element
Eq.~(\ref{eq:matrix_element_cont}) in terms of the form factors $\fpar(E_K)$
and $\fperp(E_K)$ as
\beq
\langle K(p_K)|\mathcal{V}^\mu|B_s(p_{B_s})\rangle
=
\sqrt{2\MBs}
\left[v^\mu \fpar(E_K) + p^\mu_{\perp}\fperp(E_K)\right] .
\label{eq:lattice_ff}
\enq
The relations to the original form factors $f_+$ and $f_0$ are given by
\begin{subequations}
\begin{align}
f_+(q^2) &= 
\frac{1}{\sqrt{2M_{B_s}}}
\left[\fpar(E_K) + (\MBs - E_K) \fperp(E_K)\right],
\label{eq:fp}
\\
f_0(q^2) &= 
\frac{\sqrt{2\MBs}}{M_{B_s}^2 - M_K^2} 
\left[(\MBs - E_K)\fpar(E_K) + (E_K^2 - M_K^2)\fperp(E_K)\right].
\label{eq:f0}
\end{align}
\label{eq:fp_f0_fpar_fperp}
\end{subequations}
The kinematic constraint, Eq.~(\ref{eq:kin_constraint}), is
automatically satisfied in Eq.~(\ref{eq:fp_f0_fpar_fperp}).  

In the $B_s$ rest frame, which we use throughout the lattice-QCD calculation,
the form factors $\fpar$ and $\fperp$ are related to the temporal and spatial
components of the matrix element of the vector current $\mathcal{V}^\mu$ via
\begin{subequations}
\begin{align}
\fpar(E_K)
&=
\frac{\langle K|\mathcal{V}^0|B_s \rangle}{\sqrt{2\MBs}}, \\
\fperp(E_K)
&=
\frac{\langle K|\mathcal{V}^i|B_s \rangle}{\sqrt{2\MBs}}\frac{1}{p_K^i}.
\label{eq:cont_fperp}
\end{align}
\label{eq:cont_fpar_fperp}
\end{subequations}
Note that there is no summation over the superscript $i$ in
Eq.~(\ref{eq:cont_fperp}). The continuum-QCD current is related to the lattice
current operator $V^\mu$ by a multiplicative renormalization factor, i.e.,
\beq
\mathcal{V}^\mu(x) = Z_{V_\mu} V^\mu(x) .
\label{eq:currents}
\enq
The lattice current~$V^\mu$ is defined in Sec.~\ref{subsec:operators}, below.
We use a mostly nonperturbative method to compute $Z_{V_\mu}$. The details are
explained in Sec.~\ref{subsec:operators}.

The desired matrix elements (and  hence form factors) can be calculated from
suitably defined two- and three-point correlation functions:
\begin{subequations}
\begin{align}
C_2^{B_s}(t;\bm{p}_{B_s} = 0)  &= \sum_{\bm{x}} 
\langle 
\mathcal{O}_{B_s}(0,\bm{0}) 
\mathcal{O}_{B_s}^\dagger(t,\bm{x})
\rangle ,  \\
C_2^{K}(t;\bm{p}_K)  &= \sum_{\bm{x}} 
\langle 
\mathcal{O}_{K}(0,\bm{0}) 
\mathcal{O}_{K}^\dagger(t,\bm{x})
\rangle 
e^{i \bm{p}_K\cdot \bm{x}} , \\
C_{3,\mu}^{B_s \rightarrow K}(t,T;\bm{p}_K) &= \sum_{\bm{x},\bm{y}}
\langle 
\mathcal{O}_{K} (0,\bm{0}) 
V^\mu(t,\bm{y}) 
\mathcal{O}_{B_s}^\dagger(T,\bm{x}) 
\rangle
e^{i \bm{p}_K\cdot \bm{y}},
\label{subeq:cont_3pt}
\end{align}
\label{eq:cont_2pt_3pt}
\end{subequations}
where $\mathcal{O}_{B_s}$ and $\mathcal{O}_K$ are lattice interpolating
operators, which are defined in Sec.~\ref{subsec:operators}, below. Further,
$\bm{p}_K$ is the kaon spatial momentum, whose components in a finite volume
are integer multiples of~$2\pi/N_s$, where $N_s$ is the lattice spatial
dimension in lattice units.

The basic procedure for calculating the continuum form factors $f_+$ and $f_0$
in Eq.~(\ref{eq:matrix_element_cont}) in lattice QCD is the following:
\begin{enumerate}
\item
For each ensemble:
\begin{enumerate}[label=(\roman*)]
\item 
Determine the lattice $B_s$ meson masses, kaon masses and energies from the
lattice two-point correlation functions.
\item 
Determine the lattice form factors $\fpar^{\mathrm{lat}}$ and
$\fperp^{\mathrm{lat}}$ at several discrete kaon momenta $\bm{p}_K$ from the
two- and three-point correlation functions.
\item
Obtain the renormalized form factors by matching the lattice current to the
continuum as in Eq.~(\ref{eq:currents}).
\end{enumerate}
\item 
Use chiral perturbation theory together with Symanzik effective theory to
perform a combined chiral-continuum fit to the renormalized form factors and
extrapolate them to the physical quark masses and continuum (zero lattice
spacing) limits. This yields the continuum form factors $\fpar$ and $\fperp$ as
functions of the kaon recoil energy $E_K$ in the interval covered by the
simulation, roughly $0.5$~GeV $\lesssim E_K \lesssim 1$~GeV.
\item 
Construct the continuum form factors $f_+$ and $f_0$ from $\fpar$ and $\fperp$
via Eq.~(\ref{eq:fp_f0_fpar_fperp}) and employ a $z$ expansion to parametrize
their shapes and to extrapolate them from the low-recoil range to the entire
kinematically allowed region, which extends at high recoil to $q^2 = 0$. 
\end{enumerate}

%% file: sec3/subsection-3.2-actions.tex
\subsection{Actions and parameters} \label{subsec:action}

We use lattice gauge configurations with $N_f = 2+1$ flavors generated by the
MILC Collaboration~\cite{Bazavov:2009bb,Bernard:2001av,Aubin:2004wf}. These
configurations include two degenerate dynamical light quarks, acting as $u$ and
$d$ quarks, and one heavier, $s$, quark. The gluon fields are simulated with
the one-loop improved L\"{u}scher-Weisz action~\cite{Luscher:1984xn}. The $a^2$
tadpole-improved staggered action
(asqtad)~\cite{Blum:1996uf,Lepage:1997id,Lagae:1998pe,Lepage:1998vj,
Orginos:1998ue,Orginos:1999cr,Bernard:1999xx} is used for generating dynamical
light quarks ($u$, $d$, and $s$). Reference~\cite{Bazavov:2009bb} is a review
of simulations and formalism of improved staggered quarks.

The asqtad fermion action is also used for the valence $u$, $d$, and $s$
quarks.  The heavy valence bottom ($b$) quarks use the Sheikholeslami-Wohlert
(SW) Wilson-clover action~\cite{Sheikholeslami1985572} with the Fermilab
interpretation~\cite{ElKhadra:1996mp}.

Some of the parameters used to generate the configurations are listed in
Table~\ref{tab:ensembles}. Six ensembles with three different lattice spacings,
$a \approx 0.12$, 0.09, and 0.06~fm, are used. For each lattice spacing, we
have dynamical sea quarks with light-to-strange quark mass ratio
$m^\prime_l/m^\prime_h = 0.1$\footnote{In this paper, we use primed quantities
to denote the sea quarks and the unprimed for the valence quark.}. For the
intermediate lattice spacing $a \approx 0.09~\textrm{fm}$, we have three
additional values of $m^\prime_l/m^\prime_h = 0.05, 0.15$, and 0.2 to provide
results for the chiral extrapolation. The subset of ensembles used for the
analysis is based on experience from previous semileptonic form factor
analyses~\cite{Lattice:2015tia} and~\cite{Bailey:2015dka}. The tadpole factor
$u_0$ appearing in the one-loop improved L\"{u}scher-Weisz gauge action and in
the asqtad fermion action are determined from the fourth root of the average
plaquette. 
\begin{table}[tbp]
\centering
\caption{Parameters used for generating the lattice QCD gauge fields. The
columns from left to right are approximate lattice spacing $a$ in fm, the
lattice dimensions in lattice units $N_s^3 \times N_t$, the sea-quark mass
ratios $am_l^\prime/am_h^\prime$, the gauge coupling $\beta$, the tadpole
improvement factor $u_0$, the number of gauge-field configurations
$N_\text{conf}$, and the pion mass times the box linear spatial size $M_\pi L$
($L=N_sa$). The gauge-field configurations can be downloaded using the DOI
links provided in
Refs.~\cite{asqtad_C0.1ms,asqtad_F0.2ms,asqtad_F0.2ms_b,asqtad_F0.2ms_c,
asqtad_F0.15ms,asqtad_F0.1ms_a,asqtad_F0.1ms_b,asqtad_F0.05ms,
asqtad_SF0.1ms_a,asqtad_SF0.1ms_b}.}
\label{tab:ensembles}
\begin{tabular}{lcccccc}
\hline
\hline
$\approx$a~(fm)  &$N_s^3 \times N_t$ 
        & $am_l^\prime/am_h^\prime$ &$\beta$  &$u_0$  & $N_\text{conf}$   &$M_{\pi}L$\\
\hline
 0.12~\cite{asqtad_C0.1ms} & 
                  $24^3 \times 64$    & 0.0050/0.050  & 6.76  & 0.8678   & 2099 & 3.8 \\
\hline
 0.09~\cite{asqtad_F0.2ms,asqtad_F0.2ms_b,asqtad_F0.2ms_c} &
                  $28^3 \times 96$    & 0.0062/0.031  & 7.09  & 0.8782   & 1931 & 4.1 \\
 0.09~\cite{asqtad_F0.15ms} & 
                  $32^3 \times 96$    & 0.00465/0.031 & 7.085 & 0.8781   & 1015 & 4.1 \\
 0.09~\cite{asqtad_F0.1ms_a,asqtad_F0.1ms_b} &
                  $40^3 \times 96$    & 0.0031/0.031  & 7.08  & 0.8779   & 1015 & 4.2 \\
 0.09~\cite{asqtad_F0.05ms} &
                  $64^3 \times 96$    & 0.00155/0.031 & 7.075 & 0.877805 &  791 & 4.8 \\
\hline
 0.06~\cite{asqtad_SF0.1ms_a,asqtad_SF0.1ms_b} &
                  $64^3 \times 144$   & 0.0018/0.018  & 7.46  & 0.88764  &  827 & 4.3 \\
\hline
\hline
\end{tabular}
\end{table}

The parameters used in the valence quarks and in generating correlation
functions are listed in Table~\ref{tab:valence}. The valence light quarks are
degenerate with the sea quarks, i.e., $am_l = am_l^\prime$; the valence $s$
quark masses are set to our best determination of the $s$ quark mass on each
ensemble, based on all of our analysis of the asqtad ensembles. In general
$am_h < am_h^\prime$. The heavy $b$ quark Wilson fermions with SW lattice
action are controlled by the hopping parameter $\kappa$ and the clover
coefficient of the SW action $c_\mathrm{sw}$. We use $\kappa_b^\prime$ to
denote the values used in the computation. We use the tadpole-improved
tree-level value for $c_\mathrm{sw} = u_0^{-3}$, with $u_0$ listed in
Table~\ref{tab:ensembles}. The parameter $d_1$ is used for the correlation
function generation and will be explained later in Sec.~\ref{subsec:operators}.
\begin{table}[tbp]
\centering
\caption{Parameters used for generating the valence quarks. The approximate
lattice spacing $a$ and lattice dimensions $N_s^3 \times N_t$ in the first two
column identify the ensmeble. The light valence quarks $m_l$ are degenerate
with the sea quarks $m_l^\prime$. The valence $s$ quarks $m_h$ are better tuned
than the sea $s$ quarks $m_h^\prime$. The parameters $c_\mathrm{sw}$ and
$\kappa_b^\prime$ are used in the SW action for $b$ quarks. The rotation
parameter $d_1$ is used in the current.}
\label{tab:valence}
\begin{tabular}{ccccccc}
\hline
\hline
$\approx$a~(fm)   &$N_s^3 \times N_t$ & a$m_l$/a$m_h$ &$c_\mathrm{sw}$  & $\kappa_b^\prime$  & $d_1$\\
\hline
 0.12 & $24^3 \times 64$  & 0.0050/0.0336  & 1.53   & 0.0901 & 0.09332 \\
\hline
 0.09 & $28^3 \times 96$  & 0.0062/0.0247  & 1.476  & 0.0979 & 0.096765 \\
 0.09 & $32^3 \times 96$  & 0.00465/0.0247 & 1.477  & 0.0977 & 0.096708 \\
 0.09 & $40^3 \times 96$  & 0.0031/0.0247  & 1.478  & 0.0976 & 0.096688 \\
 0.09 & $64^3 \times 96$  & 0.00155/0.0247 & 1.478  & 0.0976 & 0.0967   \\
\hline
 0.06 & $64^3 \times 144$ & 0.0018/0.0177  & 1.4298 & 0.1052 & 0.0963   \\
\hline
\hline
\end{tabular}
\end{table}

Table~\ref{tab:derived} lists the parameters derived from the lattice
simulation. The relative lattice scale is set by calculating $r_1/a$ on each
ensemble, where $r_1$ is related to the force between static quarks,
$r_1^2F(r_1) = 1.0$~\cite{Sommer:1993ce,Bernard:2000gd}. A mass-independent
procedure is used to set $r_1/a$.  We use the $r_1/a$ to convert all lattice
quantities to $r_1$ units. The physical value of $r_1$ is determined from
$f_\pi$: $r_1 = 0.3117(22)~ \mathrm{fm}$~\cite{Bazavov:2009bb,Bazavov:2011aa}.
The physical value $\kappa_b$~\cite{Bailey:2014tva}, corresponding to the
physical $b$-quark mass, and the critical value $\kappa_\mathrm{crit}$,
corresponding to the zero quark masses in the SW action on each ensemble, are
also listed in Table~\ref{tab:derived}. They will be used only for correcting
the $b$-quark masses as will be discussed in Sec.~\ref{subsec:bquark}. The
Goldstone pion mass $M_\pi$ and the root-mean-square (RMS) pion mass
$M_\pi^{\textrm{RMS}}$ are listed in the last two columns of
Table~\ref{tab:derived}.
\begin{table}[tbp]
\centering
\caption{Parameters derived from the simulation. The approximate lattice
spacing $a$ in fm and the lattice dimensions in lattice units $N_s^3 \times
N_t$ are used for identifying the ensemble. Relative scales $r_1/a$ are listed
in the third column. The statistical errors on $r_1/a$ are 0.1 to 0.3\% and the
systematic errors are comparable. The physical $\kappa_b$~\cite{Bailey:2014tva}
for the SW action are listed in the fourth column, where the first error is the
statistics plus fitting error and the second one is due to the uncertainty in
the lattice spacing. The critical $\kappa_\mathrm{crit}$ for the SW action are
listed in the fifth column. The errors of $\kappa_\mathrm{crit}$ are in the
last digit.  We also list the Goldstone pion mass ($M_\pi$) and
root-mean-square (RMS) pion mass ($M_\pi^{\textrm{RMS}}$) here.}
\label{tab:derived}
\begin{tabular}{ccccccc}
\hline
\hline
$\approx$a~(fm)   &$N_s^3 \times N_t$ &$r_1/a$   &$\kappa_b$
&$\kappa_\mathrm{crit}$	& $M_\pi$ (MeV)	&$M_\pi^{\textrm{RMS}}$ (MeV)\\
\hline
0.12 & $24^3 \times 64$  & 2.73859 & 0.0868(9)(3)	& 0.14096  & 277  & 456\\
\hline
0.09 & $28^3 \times 96$  & 3.78873 & 0.0967(7)(3)	& 0.139119 & 354  & 413\\
0.09 & $32^3 \times 96$  & 3.77163 & 0.0966(7)(3)	& 0.139134 & 307  & 374\\
0.09 & $40^3 \times 96$  & 3.75459 & 0.0965(7)(3)	& 0.139173 & 249  & 329\\
0.09 & $64^3 \times 96$  & 3.73761 & 0.0964(7)(3)	& 0.13919  & 177  & 277\\
\hline
0.06 & $64^3 \times 144$ & 5.30734 & 0.1050(5)(2)	& 0.137678 & 224  & 255\\
\hline
\hline
\end{tabular}
\end{table}

%% file: sec3/subsection-3.3-currents.tex
\subsection{Interpolating operators, currents, and correlation functions}
\label{subsec:operators}

Here we specify the interpolating operators for the kaon and $B_s$ meson and
the lattice vector current needed for the correlation functions in
Eq.~(\ref{eq:cont_2pt_3pt}). For the kaon, the local pseudoscalar interpolating
operator is used
\beq
\mathcal{O}_{K}(t,\bm{x})
= \bar{\chi}(t,\bm{x}) (-1)^{t+x_1+x_2+x_3} \chi(t,\bm{x}) ,
\label{eq:ope_kaon}
\enq
where $\chi(t,\bm{x})$ is the one-component staggered fermion field.

The $B_s$ meson interpolating operator contains a $b$-quark field, simulated
with the improved Wilson action, and a light staggered field for the
$s$-quark~\cite{Bailey:2008wp,Wingate:2002fh,Kawamoto:1981hw}
\begin{subequations}
\begin{align}
\mathcal{O}_{B_s}(t,\bm{x}) &=
\sum_{\bm{y}} \bar{\psi}(t,\bm{y}) S(\bm{y},\bm{x})
\gamma_5 \Omega(t,\bm{x}) \chi(t,\bm{x}) , \\
\Omega(t,\bm{x}) &\equiv \gamma_1^{x_1} \gamma_2^{x_2} \gamma_3^{x_3}
\gamma_4^{t},
\end{align}
\label{eq:ope_bs}
\end{subequations}
where $\psi(t,\bm{y})$ is the four-component $b$-quark field, and
$S(\bm{x},\bm{y})$ is a spatial smearing function. We use two smearing
functions for the $B_s$ meson. One is the local $S(\bm{x}, \bm{y}) =
\delta(\bm{x} - \bm{y})$. The other one is the ground-state 1S wave function of
the Richardson potential~\cite{Bazavov:2011aa}.

The lattice vector current operator in Eqs.~(\ref{eq:currents})
and~(\ref{subeq:cont_3pt}) is defined as in Refs.~\cite{Bailey:2008wp}
and~\cite{Wingate:2002fh}
\begin{align}
V^\mu(x) &= \bar{\Psi}(x) \gamma^\mu \Omega(x) \chi(x),
\label{eq:lat_V}
\end{align}
where the rotated $b$-quark field $\Psi$, defined by
\beq
\Psi = (1+a d_1 \bm{\gamma} \cdot \bm{D}_{\mathrm{lat}})\psi, 
\enq
removes $O(a)$ discretization effects from the current~\cite{ElKhadra:1996mp}.
Here $\bm{D}_{\mathrm{lat}}$ is a symmetric nearest-neighbor covariant
difference operator. The coefficient $d_1$, shown in Table~\ref{tab:valence},
is set to its tadpole-improved tree-level value so that the lattice vector
current is tree-level $O(a)$ improved. 

The renormalization constant $Z_{V_\mu}$, needed to match the lattice vector
current to its continuum counterpart (see Eq.~(\ref{eq:currents})), is
determined using a mostly nonperturbative renormalization
procedure~\cite{Harada:2001fi,ElKhadra:2001rv}:
\beq
Z_{V^\mu_{bl}} = \rho_{V^\mu} \sqrt{Z_{V^4_{bb}} Z_{V^4_{ll}}},
\label{eq:Z-factor}
\enq
where $Z_{V^4_{bb}}$ and $Z_{V^4_{ll}}$ are the renormalization factors for the
flavor-diagonal $b$- and light-quark temporal vector currents that are
calculated nonperturbatively in Ref.~\cite{Lattice:2015tia} and listed in
Table~\ref{tab:renorm-factors}. The remaining flavor-off-diagonal parameters
$rho_{V^\mu}$ are calculated to one-loop order in perturbation theory,
separately from this analysis, and also listed in
Table~\ref{tab:renorm-factors}. In order to reduce subjectivity in our
analysis, we employed a blinding procedure in the form of a small
multiplicative offset applied to the $\rho$ factors and known to only two of
the authors. This blinding factor was subsequently disclosed and removed only
after the analysis choices were finalized.
\begin{table}[tbp]
\centering
\caption{Parameters for the renormalization of the form factors. The
approximate lattice spacing $a$ and lattice dimensions $N_s^3 \times N_t$ in
the first two column identify the ensmeble. The light-light and heavy-heavy
renormalization factors $Z_{V^4_{ll}}$ and $Z_{V^4_{bb}}$ are listed in the
third and fourth columns.  The one-loop estimates of $\rho_{V^i}$ and
$\rho_{V^4}$ are listed in the fifth and sixth columns. The errors shown are
statistical. The complete current renormalization is obtained via
Eq.~(\ref{eq:Z-factor}).}
\label{tab:renorm-factors}
\begin{tabular}{cccccc}
\hline
\hline
$\approx$a~(fm) &
$N_s^3 \times N_t$  & $Z_{V^4_{ll}}$  & $Z_{V^4_{bb}}$  & $\rho_{V^i}$  & $\rho_{V^4}$\\
\hline
0.12  & $24^3 \times 64$  & 1.7410(30)  & 0.5015(8)   & 0.973082  & 1.006197 \\
\hline
0.09  & $28^3 \times 96$  & 1.7770(50)  & 0.4519(15)  & 0.975822  & 0.999308 \\
0.09  & $32^3 \times 96$  & 1.7760(50)  & 0.4530(15)  & 0.975775  & 0.999405 \\
0.09  & $40^3 \times 96$  & 1.7760(50)  & 0.4536(15)  & 0.975744  & 0.999441 \\
0.09  & $64^3 \times 96$  & 1.7760(50)  & 0.4536(15)  & 0.975703  & 0.999416 \\
\hline
0.06  & $64^3 \times 144$ & 1.8070(70)  & 0.4065(21) &  0.979176  & 0.995327 \\
\hline
\hline
\end{tabular}
\end{table}

In the generation of the correlation functions defined in
Eqs.~(\ref{eq:cont_2pt_3pt}),~(\ref{eq:ope_kaon}),~(\ref{eq:ope_bs}),
and~(\ref{eq:lat_V}), we increase statistics by repeating the calculation at
$N_\mathrm{src}$ source times evenly distributed in the $N_t$ direction. The
three-point correlation functions are generated with two adjacent temporal
source-sink separations: $T = T_\mathrm{sink}$ and $T = T_\mathrm{sink} + 1$.
Both $N_\mathrm{src}$ and $T_\mathrm{sink}$ are listed in
Table~\ref{tab:src_sink}. For the kaon recoil momenta we include the following
lowest possible values: $\bm{p}_K/ (2\pi/N_s) = (0,0,0), (1,0,0), (1,1,0), (1,
1, 1)$ and $(2,0,0)$. In practice, the largest momentum $\bm{p}_K =
2\pi(2,0,0)/N_s$ is too noisy and is excluded from the analysis.
\begin{table}[tbp]
\caption{The number of time sources $N_\mathrm{src}$ used in the two- and
three-point correlation function generation and the source-sink separations
$T_\mathrm{sink}$ used in the three-point correlation function generation. The
approximate lattice spacing $a$ and lattice dimensions $N_s^3 \times N_t$ in
the first two column identify the ensmeble.}
\label{tab:src_sink}
\centering
\begin{tabular}{cccc}
\hline
\hline
$\approx$a~(fm)   &$N_s^3 \times N_t$ &$N_\mathrm{src}$   &$T_\mathrm{sink}$\\
\hline
 0.12 & $24^3 \times 64$  & 4 & 18  \\
\hline
 0.09 & $28^3 \times 96$  & 4 & 25 \\
 0.09 & $32^3 \times 96$  & 8 & 25 \\
 0.09 & $40^3 \times 96$  & 8 & 25 \\
 0.09 & $64^3 \times 96$  & 4 & 25  \\
\hline
 0.06 & $64^3 \times 144$ & 4 & 36 \\
\hline
\hline
\end{tabular}
\end{table}

%% file: section4-analysis.tex
\section{Analysis} \label{sec:analysis}

With lattice correlation functions in hand, we follow the steps outlined near
the end of Sec.~\ref{subsec:def-formfactors} to determine the form factors
defined there, where we make use of the spectral decomposition of the
correlation functions to extract the desired parameters. The two-and
three-point functions take the form~\cite{Wingate:2002fh}:
\begin{subequations}
\begin{align}
C_2^{B_s}(t;0) &=
\sum_{n=0}^{2N-1} 
(-1)^{n(t+1)}
|Z_{B_s}^{(n)}|^2 
\left(e^{-M_{B_s}^{(n)}t} + e^{-M_{B_s}^{(n)}(N_t - t)} \right), 
\label{eq:lat_2pt_Bs} \\
C_2^K(t;\bm{p}_K) &= 
\sum_{n=0}^{2N-1} 
(-1)^{n(t+1)}
|Z_{K}^{(n)}(\bm{p}_K)|^2 
\left(e^{-E_K^{(n)}t} + e^{-E_K^{(n)}(N_t - t)} \right),
\label{eq:lat_2pt_Kaon} \\
C_{3,\mu}^{B_s \rightarrow K}(t,T;\bm{p}_K) &=
\sum_{m,n=0}^{2N-1} 
(-1)^{m(t+1)}(-1)^{n(T-t-1)} 
|Z_{B_s}^{(n)}|
|Z_{K}^{(n)}(\bm{p}_K)| 
D_{mn}^{\mu}
e^{-E_K^{(m)}t} e^{-M_{B_s}^{(n)}(T-t)},
\label{eq:lat_3pt} 
\end{align}
\label{eq:lat_2pt_3pt}
\end{subequations}
with
\begin{subequations}
\begin{align}
Z_{B_s}^{(n)} &=
\frac{|\langle 0|\mathcal{O}_{B_s}|{B_s}^{(n)} \rangle|}{\sqrt{2M_{B_s}^{(n)}}} ,
\label{eq:overlap}\\
Z_{K}^{(n)}(\bm{p}_K) &=
\frac{|\langle 0|\mathcal{O}_K|K^{(n)}(\bm{p}_K) \rangle|}{\sqrt{2E_K^{(n)}}} ,
\label{eq:overlap_K}\\
D_{mn}^{\mu} &\equiv 
\frac{
\langle K^{(m)}|V^\mu|B_s^{(n)} \rangle
}
{
\sqrt{2E_K^{(m)}} \sqrt{2M_{B_s}^{(n)}}
} .
\label{eq:Dmn_flat}
\end{align}
\label{eq:3pt_amp}
\end{subequations}
The $(-1)^{n(t+1)}$ and $(-1)^{n(T-t)}$ terms in Eq.~(\ref{eq:lat_2pt_3pt})
arise because with our choice for the light-quark valence action the
interpolating operators also generate opposite-parity (scalar) states. The
overlap factors $Z_{B_s}^{(n)}$ and $Z_{K}^{(n)}(\bm{p}_K)$ describe the
overlap of the interpolating operators with the states $|{B_s}^{(n)} \rangle$
and $|K^{(n)}(\bm{p}_K) \rangle$, respectively, while the $D_{mn}^{\mu}$
contain the desired matrix element.

In Sec.~\ref{subsec:masses}, we extract the meson masses, and the overlap
factors $Z_{B_s}^{(n)}$ and $Z_{K}^{(n)}(\bm{p}_K)$ from the two-point
correlation functions. We explain how the lattice form factors are extracted
from the two- and three-point correlation functions in
Sec.~\ref{subsec:lat_form_factors}. We briefly describe the heavy $b$-quark
mass corrections in Sec.~\ref{subsec:bquark}. The chiral-continuum fit function
and extrapolation are described in Sec.~\ref{subsec:chiral_extrap}.

\input{./sec4/subsecion-4.1-meson-masses}

\input{./sec4/subsecion-4.2-form-factors}

\input{./sec4/subsecion-4.3-b-mass-correction}

\input{./sec4/subsecion-4.4-chiral}

%% file: sec4/subsecion-4.1-meson-masses.tex
\subsection{Analysis of the two-point correlation functions}
\label{subsec:masses}

The $B_s$-meson masses, kaon masses, kaon energies, and $B_s$ and kaon overlap
factors are obtained from fitting the two-point correlation functions to the
functional forms in Eqs.~(\ref{eq:lat_2pt_Bs}) and~(\ref{eq:lat_2pt_Kaon}). 

\begin{figure}[tbp]
\centering
\includegraphics[width=0.47\columnwidth]{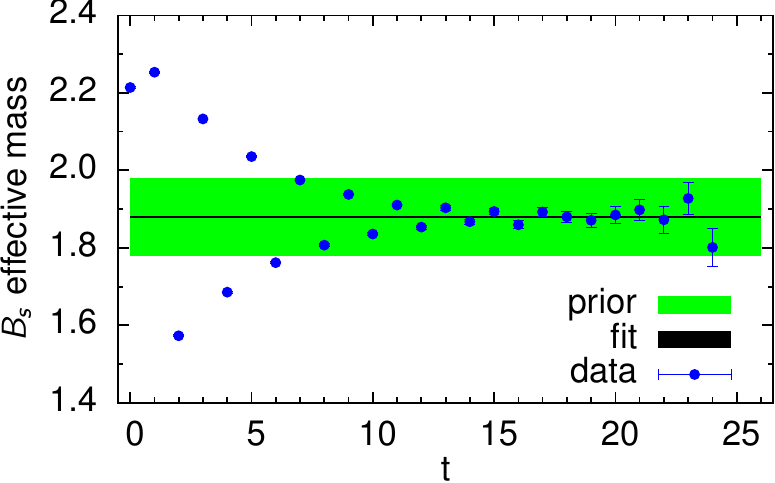}
\qquad
\includegraphics[width=0.47\columnwidth]{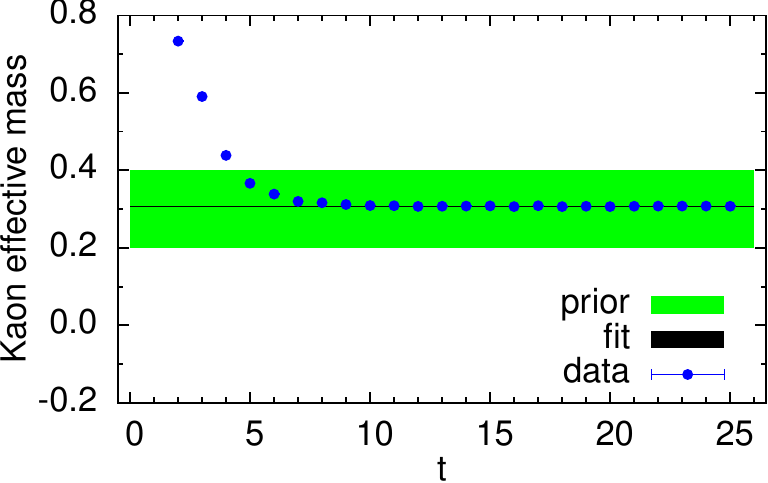}
\caption{$B_s$ and kaon meson two-point correlation function for the
$a\approx0.12$~fm, $N_s^3 \times N_t = 24^3 \times 64$ ensemble. The blue
points are the effective mass constructed via Eq.~(\ref{eq:eff_mass}). The
prior is shown as a green band. The fitted meson mass is shown as a thin gray
horizontal band. Some of the blue-point error bars are too small to be visible.
The error of the fitted kaon meson mass is magnified 20 times to make it
visible in the plot.}
\label{fig:eff_mass_Bs_Kaon}                                            
\end{figure}

As listed in Table~\ref{tab:src_sink}, there are 4 or 8 time sources for each
ensemble. The two-point correlation functions are averaged together and folded
around $N_t/2$ before constructing the ensemble-averaged propagators and
covariance matrix required for the two-point function fits. We use Bayesian
constraints with Gaussian priors to perform fits to the correlation functions
which include excited states. We vary the number of states and range of time
slices included in the fits to separate excited state contributions from the
desired ground state parameters and obtain reliable estimates of the
uncertainties.  The fit ranges are generally determined according to the
following rules: $t_\mathrm{max}$ is the largest value of $t$ where the
fractional error in the correlation function is smaller than $3\%$;
$t_\mathrm{min}$ is chosen small enough to get a good handle on the excited
states and to obtain a good correlated $p$~value as defined in
Ref.~\cite{Bazavov:2016nty}. The fit ranges $[t_\mathrm{min}, t_\mathrm{max}]$
for different lattice spacings are also adjusted so that the physical distances
are similar. Our fit functions include the same number of opposite parity
states as regular parity states. The number-of-states parameter $N$ in
Eq.~(\ref{eq:lat_2pt_3pt}) therefore refers to a fit function with the
pseudoscalar ground state plus $N-1$ of its radial excitations and $N$ scalar
states. Our central value fits have $N = 3$. The prior central values for the
ground state energies and overlap factors are guided by the effective mass and
effective amplitude evaluated at large times $t$. The effective mass
$m_\mathrm{eff}$ and effective amplitude $Z_\mathrm{eff}$ are constructed from
the two-point correlation functions via
\beq
m_\mathrm{eff} \equiv - \log\left[C_2(t+1)/C_2(t)\right], \quad
Z_\mathrm{eff}^2
\equiv e^{+ m_{\mathrm{eff}} t}  C_2(t) \,. 
\label{eq:eff_mass}
\enq
Here $C_2$ stands for the lattice two-point correlation function for the kaon
or $B_s$ meson.
The prior central values for $M_{B_s}^{(0)}$, $M_K^{(0)} \equiv E_K^{(0)}$,
$Z_{B_s}^{(0)}$, $Z_K^{(0)}(0)$, and $Z_K^{(0)}(\bm{p}_K)$ are set according to
Eq.~(\ref{eq:eff_mass}) and the widths are set to be 0.1 or larger in lattice
units. The prior central values for $M_{B_s}^{(n \neq 0)}$ and $E_K^{(n \neq
0)}$ are set using the energy difference between ground states and the
corresponding excited states from the PDG~\cite{Olive:2016xmw} values as a
guide wherever available and the widths are set to be 0.1 or larger in lattice
units. The prior central values for $Z_{B_s}^{(1)}$, $Z_{K}^{(1)}(0)$, and
$Z_{K}^{(1)}(\bm{p}_K)$ are set using the $N=1$ fit results as a guide and the
widths are set to be 0.1 or larger in lattice units. The prior central values
for $Z_{B_s}^{(2)}$, $Z_K^{(2)}(0)$, and $Z_K^{(2)}(\bm{p}_K)$ are set using
$Z_{B_s}^{(0)}$, $Z_K^{(0)}(0)$, and $Z_K^{(0)}(\bm{p}_K)$ as a guide and the
widths are set to be 0.1 or larger in lattice units. Finally the prior central
values for $Z_{B_s}^{(3,4,5)}$, $Z_K^{(3,4,5)}(0)$, and
$Z_K^{(3,4,5)}(\bm{p}_K)$ are set to be 0.1 and the widths are set to be 1.0 or
larger in lattice units. The prior widths in general are set to be large enough
so that no bias is introduced in the fits. An example of the $B_s$ effective
mass, prior, and fit result is shown in the left panel of
Fig.~\ref{fig:eff_mass_Bs_Kaon}. The corresponding kaon effective mass has
smaller oscillations and much smaller errors as shown in the right panel of
Fig.~\ref{fig:eff_mass_Bs_Kaon}. Our fit results are stable over a range of
$t_\mathrm{min}$ choices and consistent with results from $N=2$ fits. We find
that the lattice correlation functions are precise enough to determine the
first excited and opposite-parity, $N = 2$, states. Including extra $N = 3$
excited states better stabilizes the errors of fit posteriors.

\begin{figure}[tbp]
\centering
\includegraphics[width=0.47\columnwidth]{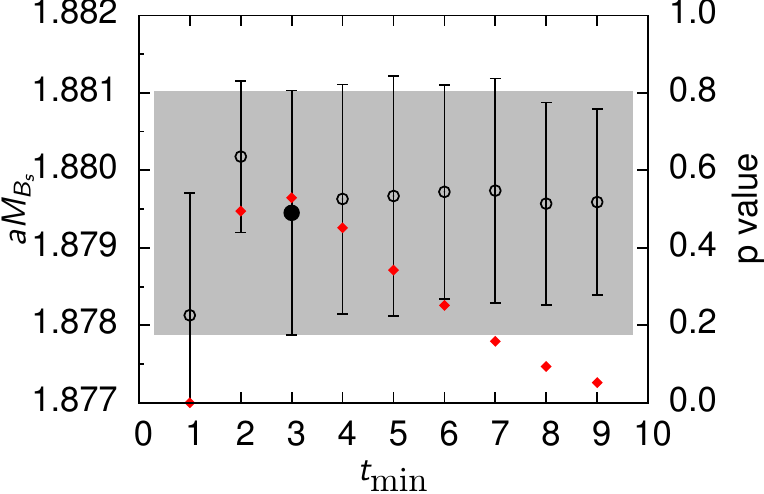}
\qquad
\includegraphics[width=0.47\columnwidth]{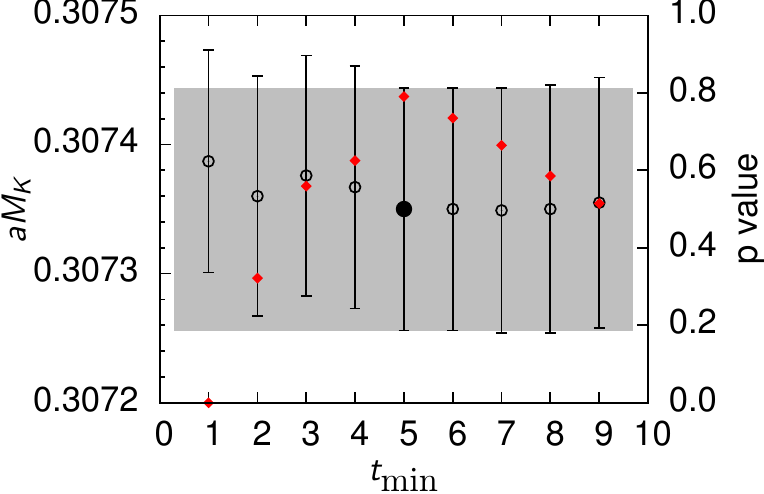}
\caption{Fitted $B_s$-meson and kaon masses, $\MBs$ and $M_K$ in lattice units
at different $t_\mathrm{min}$ for the $a\approx0.12$~fm, $N_s^3 \times N_t =
24^3 \times 64$ ensemble. The left vertical axes show the fitted masses and the
right vertical axes show the corresponding $p$~value of the fit. The chosen fit
results are shown in wide gray bands. The masses are shown in circles with
error bars. The selected value of $t_{\rm min}$ is plotted using a solid circle
and its error band is extended across the plot in gray. Red diamonds denote the
$p$ values.}
\label{fig:2pts_Bs_kaon}                                            
\end{figure}

The left panel of Fig.~\ref{fig:2pts_Bs_kaon} shows an example of the stability
plot for the $B_s$ meson. Fit intervals are chosen based on these plots and are
listed in Table~\ref{tab:fit_ranges}. Representative fit results for the kaon
are shown in the right panel of Fig.~\ref{fig:2pts_Bs_kaon}.
\begin{table}[tbp]
\centering
\caption{Fit ranges $[t_\text{min},t_\text{max}]$ used in the kaon and $B_s$
meson two-point correlator fits.}
\label{tab:fit_ranges}
\begin{tabular}{ccc}
\hline
\hline
$\approx$a~(fm)   &Kaon &$B_s$ meson \\
\hline
0.12    &[5,31] &[3,22] \\
0.09    &[7,47] &[4,30] \\
0.06    &[10,71] &[6,44] \\
\hline
\hline
\end{tabular}
\end{table}

\begin{figure}[tbp]
\centering
\includegraphics[width=0.47\columnwidth]{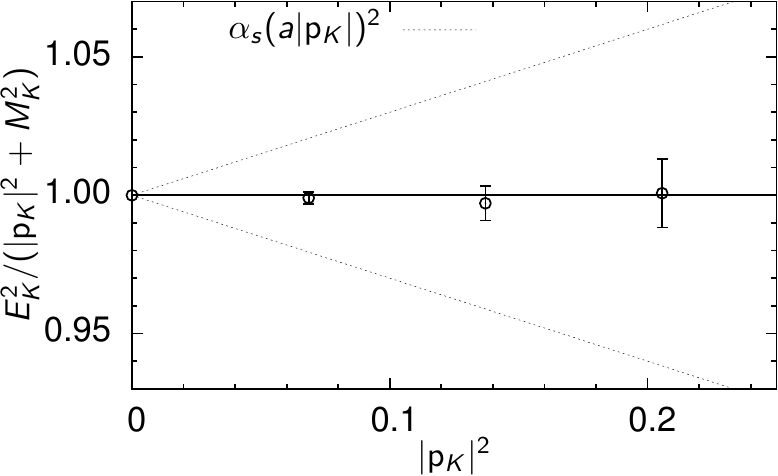}   
\qquad
\includegraphics[width=0.47\columnwidth]{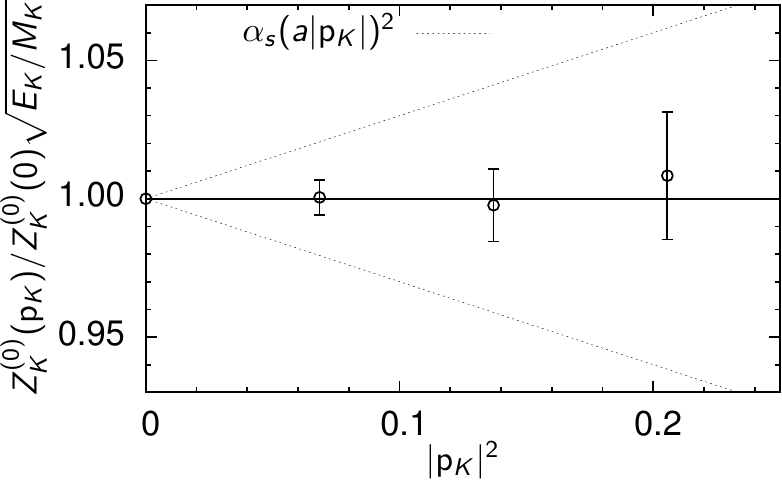}   
\caption{Test of Eq.~(\ref{eq:disp}) for the $a\approx0.12$~fm, $N_s^3 \times
N_t = 24^3\times 64$ ensemble. Left: energy-momentum dispersion relation where
$E_K$ and $M_K$ come from the kaon 2-point correlators. Right: test of wave
function overlap momentum dependence. The dashed lines on both plots show the
power-counting estimate of the size of the momentum-dependent discretization
error, $\mathcal{O}(\alpha_s|\bm{p}_K|^2a^2)$.}
\label{fig:dispersion}                                            
\end{figure}

The fit results for the kaon energies and overlap factors can be compared with
the continuum relations
\beq
E_K^2 = M_K^2 + \bm{p}_K^2,
\quad
Z_K^{(0)}(\bm{p}_K) = Z_K^{(0)}(0) \sqrt{\frac{M_K}{E_K}}
\label{eq:disp}
\enq
to study momentum-dependent discretization errors. As illustrated in
Fig.~\ref{fig:dispersion} we find that the $E_K$ and $Z_K^{(0)}(\bm{p}_K)$
satisfy Eq.~(\ref{eq:disp}) albeit with increasing statistical errors at higher
momenta. We therefore use the continuum relations for the kaon energies and
$Z_K$ factors whenever possible. 

%% file: sec4/subsecion-4.2-form-factors.tex
\subsection{Extracting form factors from two- and three-point correlation
functions} \label{subsec:lat_form_factors}

The form factors are related to the semileptonic matrix elements via
Eq.~(\ref{eq:cont_fpar_fperp}), and the lattice matrix elements are contained
in the three-point correlation function as in Eqs.~(\ref{eq:lat_3pt})
and~(\ref{eq:3pt_amp}). To get the lattice form factors $f_{\parallel,
\perp}^{\textrm{lat}}$, we fit the two- and three-point correlation functions
together. In particular, we perform combined two- and three-point
correlation-function fits according to Eqs.~(\ref{eq:lat_2pt_3pt})
and~(\ref{eq:3pt_amp}) with $N=3$. The three-point fit ranges are chosen to be
$[t_\text{min}^K,T - t_\text{min}^{B_s}]$ with $T=T_\mathrm{sink}$ or
$T_\mathrm{sink}+1$. The parameters to be fitted are $M_{B_s}^{(n)}$,
$E_K^{(n)}$, $Z_{B_s}^{(n)}$, $Z_K^{(n)}(\bm{p}_K)$, and $D_{mn}^{\mu}$. The
prior central values for the $B_s$-meson and kaon masses, $Z_{B_s}^{(0)}$, and
$Z_K^{(0)}(\bm{p}_K)$ are chosen as the posteriors of the two-point correlator
fits. The kaon energies and $Z_K^{(0)}(\bm{p}_K)$ are constrained according to
Eq.~(\ref{eq:disp}). The $Z_{B_s}^{(n \neq 0)}$ and $Z_K^{(n\neq 0)}(\bm{p}_K)$
central values are taken to be the same size as $Z_{B_s}^{(0)}$ and
$Z_K^{(n\neq 0)}(0)$. The priors for $D_{00}^\mu$ are guided by the constructed
ratio $\bar{R}_{3,0}^{B_s\to K}(t,T)$ defined in Refs.~\cite{Lattice:2015tia}
and~\cite{Bailey:2015dka}. The prior widths for the above parameters are chosen
to be 0.1 or larger in lattice units. The priors for all the other $D_{mn}^\mu$
are chosen to be $0.1\pm2.0$. The ground-state energies obtained from the
combined two- and three-point correlator fits are consistent with those from
the two-point fits as described in Sec.~\ref{subsec:masses}.

\begin{figure}[tbp]
\centering
\includegraphics[width=0.47\columnwidth]{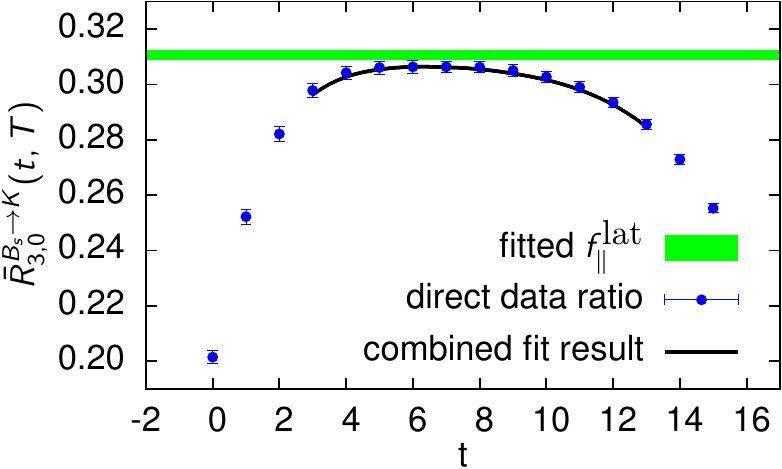}
\caption{The lattice form factor from a combined two-point and three-point
correlation function fit for the $a\approx0.12$~fm, $N_s^3 \times N_t = 24^3
\times 64$ ensemble. The green band is the combined fit result for
${\fpar}^{\textrm{lat}}$. The blue points with errors are obtained from the
ratio defined in Refs.~\cite{Lattice:2015tia} and~\cite{Bailey:2015dka}. The
black curve is the ratio constructed directly from combined fit results. The
small difference between the green band and the ratio comes from excited state
contributions still present in the ratio but accounted for in the fit method
used here.}
\label{fig:ratio_3pt}                                            
\end{figure}

\begin{figure}[tbp]
\centering
\includegraphics[width=0.47\columnwidth]{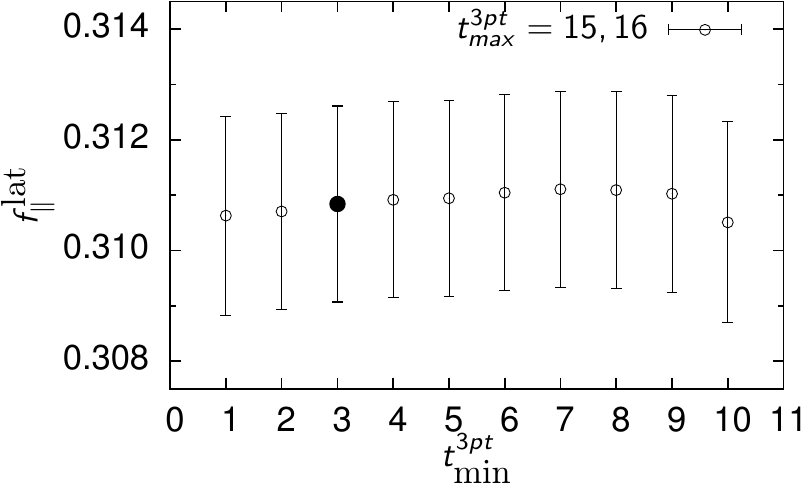}
\qquad
\includegraphics[width=0.47\columnwidth]{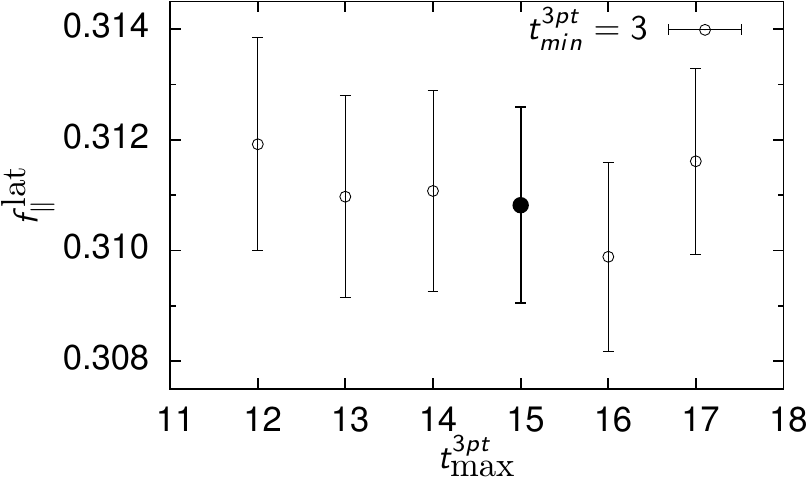}
\caption{Fit results of $\langle K^{(0)}|V^0|B_s^{(0)} \rangle$ from different
fit ranges for the $a\approx0.12$~fm, $N_s^3\times N_t = 24^3 \times 64$
ensemble with lattice kaon momentum $\bm{p}_K = (2\pi/N_s)(0,0,0)$. Left: The
three-point correlator fit maximum are fixed to be $t_{\rm{max}}^{\rm{3pt}} =
$15 and 16 and the minimum are varied between 1 and 10. Right: The three-point
correlator fit minimum are fixed to be $t_{\rm{min}}^{\rm{3pt}} = 3$ and the
maximum is varied between 12 and 17. The preferred fit ranges are shown with
filled points.}
\label{fig:ratio_3pt_stability}
\end{figure}

Figure~\ref{fig:ratio_3pt} shows that the fitted ${\fpar}^{\textrm{lat}}$
coming from the combined fit is in slight tension with the constructed ratio
$\bar{R}_{3,0}^{B_s\to K}(t,T)$ defined in Refs.~\cite{Lattice:2015tia}
and~\cite{Bailey:2015dka}. This small difference comes from excited state
contributions still present in the ratio but accounted for in the fit method
used here. We find that they are significant at the present level of precision.
Figure~\ref{fig:ratio_3pt_stability} shows an example of the stability of the
fit result when varying the fit range.

In summary, the form factors $\fpar$ and $\fperp$ are obtained from
$D_{00}^\mu$ and $E_K$ according to Eqs.~(\ref{eq:cont_fpar_fperp})
and~(\ref{eq:Dmn_flat}), after adding the renormalization factors as in
Eq.~(\ref{eq:cont_fpar_fperp}).

%% file: sec4/subsecion-4.3-b-mass-correction.tex
\subsection{Heavy bottom quark mass correction} \label{subsec:bquark}

The heavy valence $b$ quark is simulated with the Sheikholeslami-Wohlert (SW)
action~\cite{Sheikholeslami1985572} with the Fermilab
interpretation~\cite{ElKhadra:1996mp}. The $b$-quark mass is controlled by the
hopping parameter $\kappa_b$. The hopping parameter $\kappa_b^\prime$ used in
the simulations differs slightly from the physical value $\kappa_b$ as can be
seen in Tables~\ref{tab:valence} and~\ref{tab:derived}. We need to correct the
form factors to account for these small shifts. A detailed description of the
$\kappa$ tuning analysis and results is provided in Appendix C of
Ref.~\cite{Bailey:2014tva}. We use the method described in
Refs.~\cite{Lattice:2015tia} and~\cite{Bailey:2015dka} to adjust the form
factors to account for the slightly mistuned values of~$\kappa_b$. The relative
change in the form factors under small variations of the $b$-quark mass can be
described as
\beq
f(m_2^\prime) = f(m_2)
\left[
1 - \frac{\partial \ln f}{\partial \ln m_2}
\left(
\frac{m_2}{m_2^\prime} - 1
\right)
\right],
\enq
where $m_2$ is the physical $b$-quark kinetic mass, $m_2^\prime$ is the
$b$-quark mass used in the production run. The slopes $\frac{\partial \ln
f}{\partial \ln m_2}$ were determined in Ref.~\cite{Lattice:2015tia}. The
corrections to the form factors are about 0.1--1.8\% on different ensembles.

%% file: sec4/subsecion-4.4-chiral.tex
\subsection{Chiral-continuum extrapolation} \label{subsec:chiral_extrap}

The lattice form factors extracted from the correlation functions as described
in Sec.~\ref{subsec:lat_form_factors} are obtained  at the three finite lattice
spacings and unphysical light-quark masses listed in Table~\ref{tab:ensembles}.
Here we extrapolate them to the continuum limit and physical light-quark masses
using SU(2) hard-kaon heavy-meson rooted staggered chiral perturbation theory
(HMrS$\chi$PT)~\cite{Aubin:2005aq,Aubin:2007mc}. Based on previous
experience with the analyses of similar processes in
Refs.~\cite{Lattice:2015tia} and~\cite{Bailey:2015dka}, this best describes the
data. Heavy-quark discretization effects are also taken into account in the
chiral-continuum extrapolation. 

We employ the HMrS$\chi$PT expansion at next-to-leading order (NLO) in SU(2),
leading order in $1/M_B$, where $M_B$ is the $B$-meson mass, and include
next-to-next-to-leading-order (NNLO) analytic and generic discretization terms.
In the SU(2) hard-kaon limit, the valence and sea $s$-quark masses are taken to
be infinitely heavy and hence dropped from the HMrS$\chi$PT formula; the large
kaon energy is integrated out, and its effects are absorbed into the low-energy
constants (LECs). In Ref.~\cite{Bailey:2015dka}, the conversion rules for $B
\rightarrow K$ and $B \rightarrow \pi$ processes from SU(3) HMrS$\chi$PT to
SU(2) hard-kaon and hard-pion limits were derived. Here we follow the same
procedure to obtain the corresponding formula for $\bsk$. The details are
presented in Appendix~\ref{app:su2-chiral}. 

The NLO expression for $\bsk$ form factors in the SU(2) hard-kaon limit that we
obtain is 
\beq
f_{P, \mathrm{NLO}} =  f_{P}^{(0)} 
\left[
c^{0}_{P}(1+
\delta f_{P, \mathrm{logs}}^{\mathrm{SU(2)}}
)+ c^{\text{1}}_{P}\chi_\text{l}+
c^{\text{2}}_{P}\chi_\text{h}+ c^{3}_{P}{\chi_E}+ c^{4}_{P}\chi_E^{2}+
c^5_{P}{\chi_a^{2}} 
\right], 
\label{eq:chiral_f_nlo} 
\enq 
where $P =~\parallel$ or $\perp$, $\delta f_{P,
\mathrm{logs}}^{\mathrm{SU(2)}}$ are the non-analytic contributions from the
light-quark mass and lattice spacing, and the $\chi$ variables are
dimensionless. They are defined in Eqs.~(\ref{eq:logs_su2})
and~(\ref{eq:chi_lhEa2}). The leading-order factor is
\beq f_P^{(0)} =
\frac{1}{f_\pi}\frac{g_\pi}{E_K + \Delta_P^*}, 
\label{eq:chiral_leading}
\enq 
where $f_\pi$ is the decay constant involved, and $g_\pi$ is the $B^*B\pi$
coupling constant\footnote{$SU(3)$ breaking effects renormalize the
$g_\pi/f_\pi$ ratio; however, since it results in a overall multiplicative
factor, it has been reabsorbed in the fitting coefficients.}. The $\Delta_P^*$
term takes the pole contribution into account and is determined by requiring
$\fpar$ and $\fperp$ to have the same poles as the physical form factors $f_0$
and $f_+$, respectively. This is reasonable because, by
Eq.~(\ref{eq:fp_f0_fpar_fperp}), $\fpar$ is dominated by the $0^+$
contributions of $f_0$, and $f_\perp$, by the $1^-$ contributions of $f_+$ in
the $q^2$ range considered. Using Eq.~(\ref{eq:qsq_E}), one obtains the exact
expression for $\Delta_P^*$:
\beq
\Delta_P^* = \frac{M_{B^*}^2 - M_{B_s}^2 - M_K^2}{2M_{B_s}}.
\label{eq:Delta_P}
\enq
The vector meson (with $J^P = 1^-$) has been experimentally
measured~\cite{Tanabashi:2018oca} to be $M_{B^*} = 5324.65(25)~\mathrm{MeV}$;
the scalar $B^*$ meson (with $J^P = 0^+$) has not been observed experimentally,
but a lattice calculation~\cite{Gregory:2010gm} estimates the mass difference
between $0^+$ and $0^-$ states to be around $400~\mathrm{MeV}$: 
\beq
M_{B^*}(0^+) - M_B \approx 400~\mathrm{MeV}.
\label{eq:scalar_pole}
\enq
The vector-meson mass $M_{B^*}$ is below the $B\pi$ production threshold that
is involved in the $\bsk$ decay, and the scalar-meson mass $M_{B^*}(0^+)$ is
above the threshold. The inclusion of the scalar pole and its exact location
have little impact on the chiral fit results but stabilizes the form factor
extrapolations.

\begin{figure}[tbp]
\centering
\includegraphics[width=0.45\columnwidth]{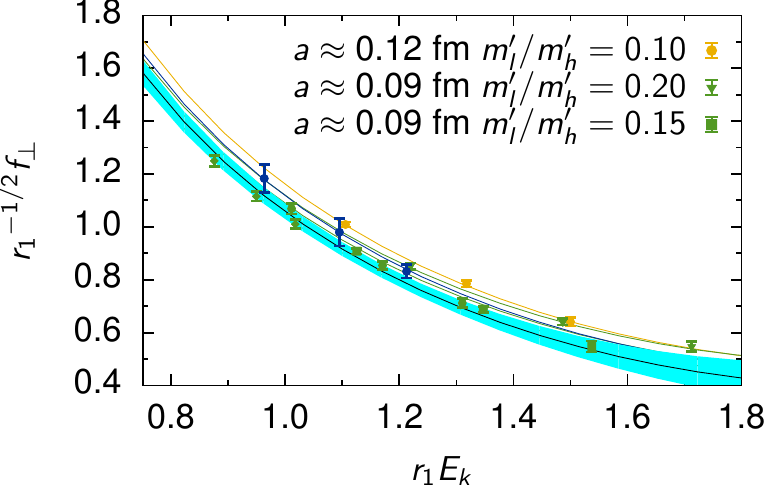}
\qquad
\includegraphics[width=0.45\columnwidth]{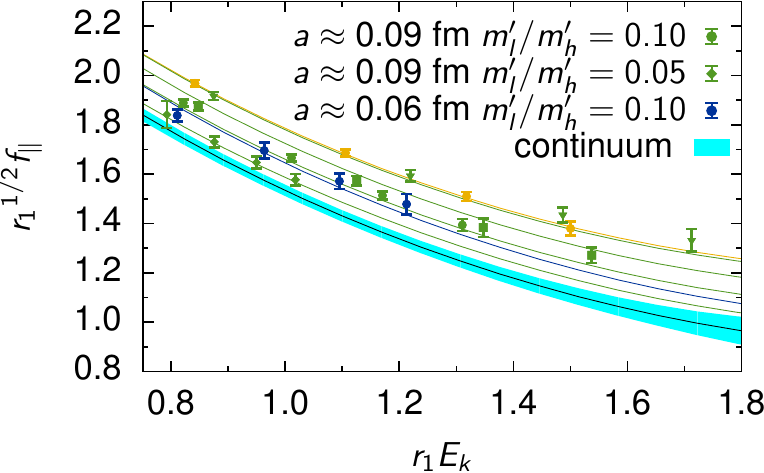}
\caption{Chiral-continuum extrapolated form factors $\fpar$ and $\fperp$ in
$r_1$ units as functions of the recoil energy $r_1E_K$. The color denotes the
lattice spacings and the symbols denote the ratio of the sea-quark masses
$m_l^\prime/m_h^\prime$. The colored fit lines correspond to the fit results
evaluated at the parameters of the ensembles. The cyan band with the black
curve shows the chiral-continuum extrapolated results.}
\label{fig:fpar_fperp_NNLO}
\end{figure}

NNLO analytic terms are included in the fits to take into account higher-order
contributions. The leading heavy $b$-quark discretization effects are also
included. The expressions for the NNLO fit functions are 
\begin{subequations}
\begin{align}
f_{P, \mathrm{NNLO}} &\equiv
f_{P, \mathrm{NNLO+HQ}} =
\left(
f_{P, \mathrm{NLO}} + f_P^{(0)} \delta f_\mathrm{NNLO}
\right)
\times (1+ \delta f_{HQ}),\\
\delta f_\mathrm{NNLO} &= 
c_P^6\chi_{l}\chi_E + 
c_P^7\chi_{a^2} \chi_E + 
c_P^8\chi_{E}^3 + 
c_P^9\chi_{l}^2 + 
c_P^{10}\chi_{l}\chi_{E}^2 
\nonumber \\
&+
c_P^{11}\chi_{a^2} \chi_{l} + 
c_P^{12}\chi_{a^2} \chi_{E}^2 + 
c_P^{13}\chi_{a^2}^2 + 
c_P^{14}\chi_{E}^4, 
\label{eq:fun_del_NNLO}
\\
\delta f_\mathrm{HQ} &=
(h_P^1 f_E + h_P^2 f_X + h_P^3 f_Y)(a\Lambda)^2 \nonumber \\
&
+(h_P^4 f_B + h_P^5 f_3)(\alpha_s a\Lambda) + h_P^6 \alpha_s(a\Lambda)^2 ,
\label{eq:fun_HQ}
\end{align}
\label{eq:fun_NNLO}
\end{subequations}
where the heavy-quark discretization effects are modeled with $\delta
f_\mathrm{HQ}$. The mismatch functions $f_{E, X, Y, B, 3}$ are defined in the
appendix of Ref.~\cite{Bazavov:2011aa}. The next-to-leading-order (NLO)
analytic term $f_{P, \mathrm{NLO}}$ was defined previously in
Eq.~(\ref{eq:chiral_f_nlo}). A Bayesian method is used in the chiral-continuum
fit. The priors are listed in Table~\ref{tab:chiral_prior}. The fit results
using the NNLO fit function in Eq.~(\ref{eq:fun_NNLO}) are used as the central
fit and are shown in Fig.~\ref{fig:fpar_fperp_NNLO}.
\begin{table}[tbp]
\caption{Priors used in the chiral-continuum extrapolation fit.
$c_p^\mathrm{NLO}$ represents $c_P^0, \cdots, c_P^5$ as shown in
Eq.~(\ref{eq:chiral_f_nlo}).  $c_p^\mathrm{NNLO}$ represents $c_P^6, \cdots,
c_P^{14}$ as shown in Eq.~(\ref{eq:fun_del_NNLO}). $h_P$ represents $h_P^1,
\cdots, h_P^6$ as appears in Eq.~(\ref{eq:fun_HQ}).}
\label{tab:chiral_prior}
\begin{tabular}{ccccc} 
\hline
\hline 
$f_\pi$ & $g_\pi$   & $c_P^\mathrm{NLO}$    & $c_P^\mathrm{NNLO}$ &
 $h_P$\\
\hline
130.4 MeV   & 0.45(8)   &0(1.0) & 0(0.6) & 0(1.0)\\
\hline
\hline  
\end{tabular} 
\end{table}

%% file: section5-errors.tex
\section{Systematic error estimations} \label{sec:errors}

The chiral-continuum extrapolated form factors are given in
Sec.~\ref{sec:analysis}. The statistical-fit errors, which are propagated
through each step of the analysis already include the effects of NNLO terms in
the chiral expansion as well as light- and heavy-quark discretization. Here we
discuss tests of the robustness of this error estimate to check for the
presence of residual truncation effects. We also consider other sources of
error not already included in our chiral-continuum fit function and construct a
complete systematic error budget over the range of $q^2$ for which we have
lattice data, $19~{\rm GeV} ^2 \lesssim q^2 \lesssim 24~{\rm GeV}^2$.
 
\input{./sec5/subsection-5.1-chiral}

\input{./sec5/subsection-5.2-current}

\input{./sec5/subsection-5.3-scale}

\input{./sec5/subsection-5.4-qmass}

\input{./sec5/subsection-5.5-bmass}

\input{./sec5/subsection-5.6-volume}

\input{./sec5/subsection-5.7-budget}

%% file: sec5/subsection-5.1-chiral.tex
\subsection{Chiral-continuum extrapolation errors} \label{subsec:chiral_err}

Our central fit uses the NNLO SU(2) hard kaon HMrS$\chi$PT fit function
described in Eq.~(\ref{eq:fun_NNLO}). In order to study truncation effects, we
consider variations of the central fit function. We also perform fits which
include fewer form factor data.

\begin{figure}[tbp]
\centering
\includegraphics[width=0.45\columnwidth]{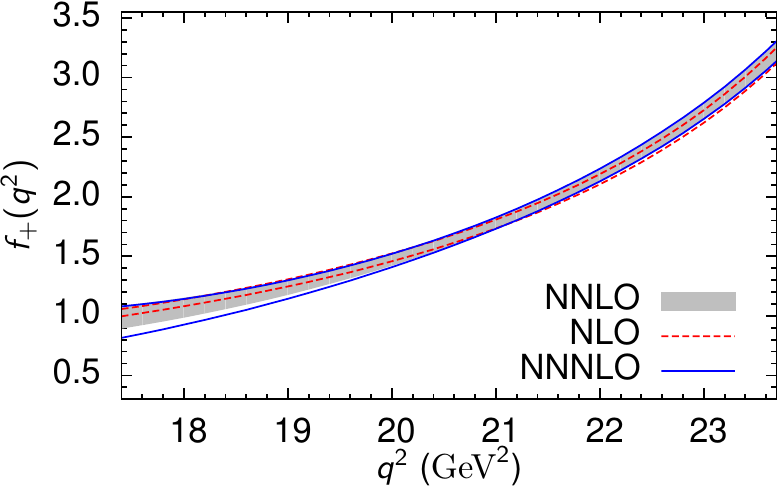}
\caption{Comparison among chiral-continuum extrapolated results for $f_+$ with
different analytic terms. The gray band shows the preferred fitting result with
NNLO SU(2) HMrS$\chi$PT. The red (dashed) and blue (solid) curves show the
error ranges resulting from the fits with only NLO analytic terms and with all
terms up to NNNLO, respectively.}
\label{fig:chiral_NLO_NNLO_NNNLO}
\end{figure}

We estimate chiral truncation effects by comparing our central NNLO fit with
fits using either only the NLO function,  defined in
Eq.~(\ref{eq:chiral_f_nlo}), or a fit function that includes the complete set
of next-to-NNLO (NNNLO) terms. The coefficients of the NNNLO terms are
constrained with the same priors as the NNLO ones.
Figure~\ref{fig:chiral_NLO_NNLO_NNNLO} shows the comparison of results for the
$f_+$ form factor from the three fits. The corresponding results for $f_0$ are
similar. We see that the results from the three different fits are consistent
with each other over the range of $q^2$ where the simulation data are located.
The NNNLO errors at small $q^2$ are larger since the data points in that region
are scarce as can be seen in Fig.~\ref{fig:fpar_fperp_NNLO}, and the fits
cannot determine the higher order terms accurately. The truncation errors are
well saturated in the $q^2 \gtrsim 19~\mathrm{GeV}^2$ region, and therefore it
is unnecessary to add an additional systematic error.

\begin{figure}[tbp]
\centering
\includegraphics[width=0.45\columnwidth]{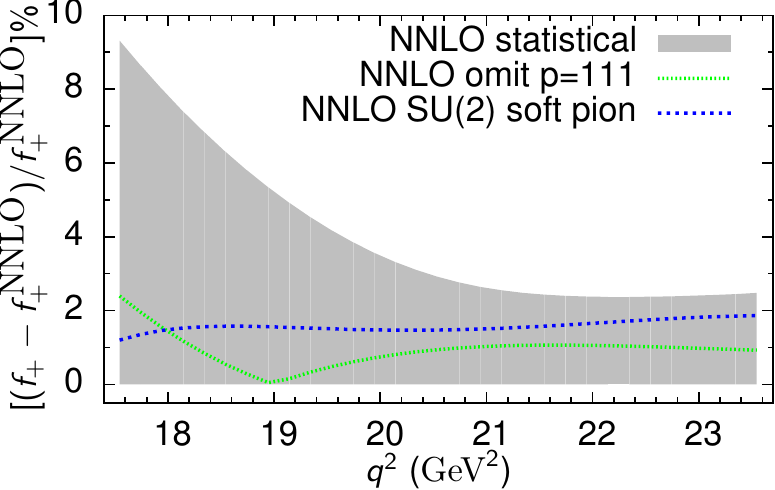}
\qquad
\includegraphics[width=0.45\columnwidth]{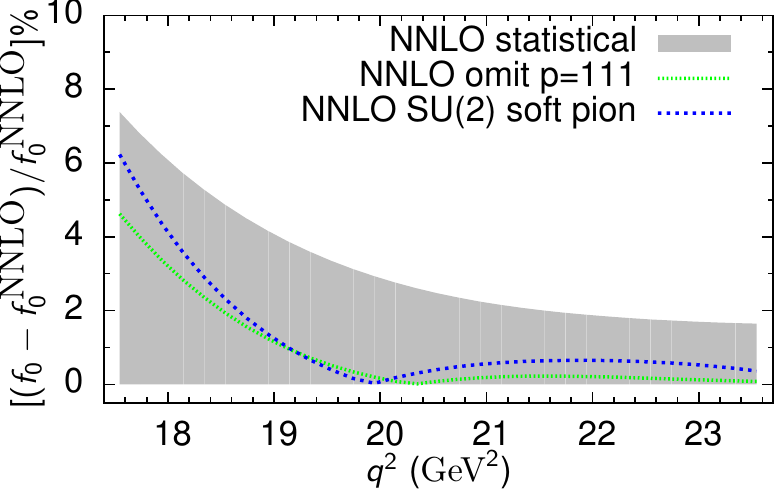}
\caption{Percent deviations of alternative chiral-continuum extrapolations from
the preferred central fit of $f_+$ and $f_0$. The curves show the deviation
from the preferred central fit obtained by either omitting the $\bm{p}_K = 2\pi
(1,1,1)/N_s$ data points or by using SU(2) soft pion HMrS$\chi$PT formula. The
gray band shows the statistical errors from the preferred NNLO SU(2)
HMrS$\chi$PT fits. The deviations are smaller than the statistical errors.}
\label{fig:sys_errors}
\end{figure}

The SU(2) hard-kaon formula is used for the central fit. To see how other
choices of the HMrS$\chi$PT formula affect the fits, we performed the fit with
soft-kaon HMrS$\chi$PT. The resulting difference is small, especially for the
$f_+$ form factor. This can be seen in Fig.~\ref{fig:sys_errors}. Since the
valence $s$-quark masses are not equal to the sea ones, the corresponding SU(3)
HMrS$\chi$PT formula are extremely complicated. We therefore did not perform
any trial fits with the SU(3) formula. Nevertheless, from previous
experience,~\cite{Lattice:2015tia, Bailey:2015dka}, SU(3) HMrS$\chi$PT
typically does not provide a good description of the data.

Our results with kaon momentum up to $2\pi(1,1,1)/N_s$ are used in the central
chiral fit. To check how the kaon energy range affects the results, we perform
the fit omitting the $\bm{p}_K = 2\pi(1,1,1)/N_s$ data. The differences are
shown in Fig.~\ref{fig:sys_errors}. Again, the difference is small especially
for the $f_+$ form factor and also for $f_0$ at $q^2 \gtrsim
19~\mathrm{GeV}^2$. 

Based on the tests discussed above and visually summarized in
Figs.~\ref{fig:chiral_NLO_NNLO_NNNLO} and \ref{fig:sys_errors}, we find that
the deviations between the results from the central fit and the alternative
fits are smaller than the statistical error of the preferred central fit. We
therefore do not assign additional systematic errors due to these sources.

%% file: sec5/subsection-5.2-current.tex
\subsection{Current renormalization uncertainties} \label{subsec:current_err}

The mostly nonperturbative renormalization procedure, described in
Eq.~(\ref{eq:currents}), used to renormalize the matrix elements (and hence the
form factors) requires, as inputs, the factors $Z_{V^4_{bb}}$, $Z_{V^4_{ll}}$,
$\rho_{V^4}$ and $\rho_{V^i}$. We estimate the error on $\fpar,\fperp$ due to
the uncertainties of the nonperturbatively determined $Z_{V^4_{bb}}$ and
$Z_{V^4_{ll}}$ by varying their central values by one standard deviation in
each direction. As expected, the resulting changes in the form factors are
small, yielding errors on $\fpar,\fperp$ in the range of 0.2--0.3\%.

For $\rho_{V^4}$ and $\rho_{V^i}$, the dominant source of error is the
truncation at one-loop order in perturbation theory. As seen in
Table~\ref{tab:renorm-factors}, the one-loop corrections provided by
$\rho_{V^4}$ and $\rho_{V^i}$ are small, with $\rho_{V^4}$ ($\rho_{V^i}$)
deviating from unity by less than $1\%$ ($2.4\%$). Here, we adopt the estimate
of the perturbative truncation error presented in Ref.~\cite{Lattice:2015tia},
which yields an uncertainty of $1\%$ on both $\rho_{V^4}$ and $\rho_{V^i}$.
This estimate is consistent with the observed differences between
nonperturbative~\cite{Chakraborty:2014zma} and
perturbative~\cite{ElKhadra:2007qe} calculations of $\rho_{V^4}=\rho_{A^4}$,
discussed in Ref.~\cite{Chakraborty:2014zma}. In particular, the observed
differences decrease in the continuum limit, as expected. Note, however, that
the nonperturbative result~\cite{Chakraborty:2014zma} employs the HISQ action
for the light quarks, while our one-loop results~\cite{ElKhadra:2007qe} employ
the asqtad action. For this reason, the comparison is suggestive but not
definitive.

%% file: sec5/subsection-5.3-scale.tex
\subsection{Lattice-scale uncertainties} \label{subsec:scale_err}

The dimensionful form factors $\fperp$ and $\fpar$, and meson energies and
masses are converted to physical units via the relative scales $r_1/a$ listed
in Table~\ref{tab:derived} and the absolute scale  $r_1 = 0.3117(22)$
fm~\cite{Bazavov:2011aa}. The statistical errors on $r_1/a$ are small and their
effects on the form factors can be neglected. We estimate the error due to the
uncertainty of $r_1$, as before, by shifting its value by one standard
deviation and repeating the chiral fit. The shifts on the form factors
$f_{+,0}$ are at most $0.8\%$ in the range of simulated momenta.

%% file: sec5/subsection-5.4-qmass.tex
\subsection{Quark mass uncertainties} \label{subsec:mass_err}

The continuum physical form factors are obtained by evaluating the
chiral-continuum extrapolated functions, as discussed in
Sec.~\ref{subsec:chiral_extrap} at the physical averaged $u$- and $d$-quark
masses, namely  $r_1 m_{ud}^\text{phys} = 0.000965(33)$, and the physical
$s$-quark mass $r_1 m_{s}^\text{phys} = 0.0265(8)$ as determined by analyzing
the light pseudoscalar meson spectrum~\cite{Bazavov:2009bb}. The error due to
the uncertainties in these masses is obtained by varying their central values
by one standard deviation to find the corresponding changes in the form
factors. The maximum changes are below $0.15\%$ in the simulated $q^2$ region.

%% file: sec5/subsection-5.5-bmass.tex
\subsection{Uncertainties arising from the bottom quark mass correction}
\label{subsec:b-mass-err}

As explained in Sec.~\ref{subsec:bquark}, the form factors are adjusted to
account for the slightly mistuned valence $b$-quark masses before the
chiral-continuum extrapolation. This accounts for the dominant effect from
$b$-quark mass mistuning. The errors on the form factors due to the
uncertainties in the $\kappa_b$-correction factors and the tuned $\kappa_b$
values are taken into account by following the procedure described in
Ref.~\cite{Lattice:2015tia}. A $q^2$-independent $0.4\%$ error due to tuning
$\kappa_b$ is assigned to both $f_+$ and~$f_0$.

%% file: sec5/subsection-5.6-volume.tex
\subsection{Finite volume effects} \label{subsec:vol_err}

Finite-volume effects, estimated by comparing infinite-volume integrals with
finite sums in HMrS$\chi$PT, are negligibly
small~\cite{Lattice:2015tia,Bailey:2015dka}, so they are omitted from the total
error budget.

%% file: sec5/subsection-5.7-budget.tex
\subsection{Summary of the statistical and systematic error budgets}
\label{subsec:chiral_summary}

\begin{figure}[tbp]
\centering
\includegraphics[width=0.45\columnwidth]{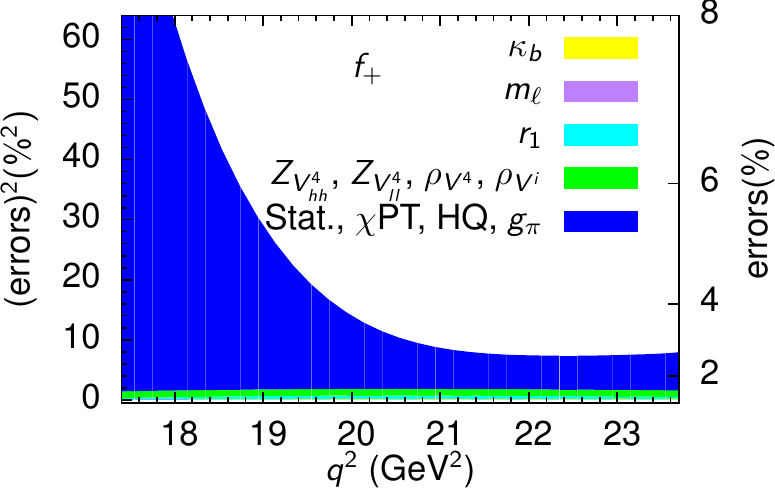}
\qquad
\includegraphics[width=0.45\columnwidth]{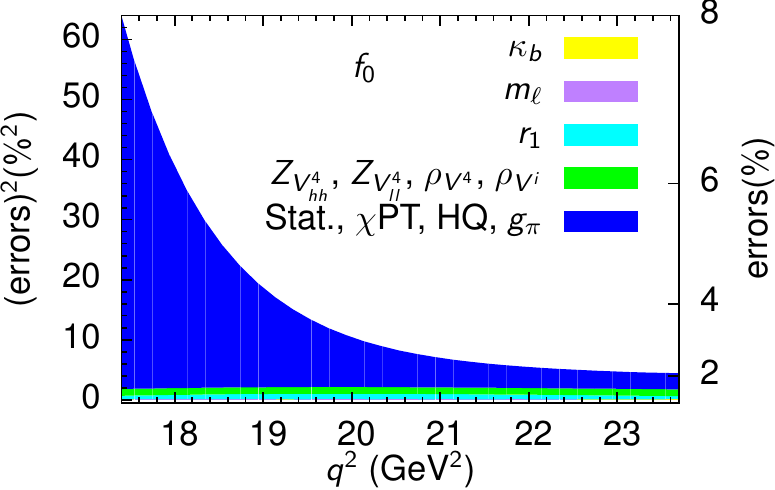}
\caption{Distribution of the errors for $f_+$ (left) and $f_0$ (right) as a
function of $q^2$. The left y axis shows the square of the errors added in
quadrature. The right y axis shows the errors themselves. The different bands
show the total error when adding individual source of error in quadrature one
by one. The error bands associated with $\kappa_b$ and $m_\ell$ are too small
to be visible on the plots.}
\label{fig:error_budgets}
\end{figure}

The systematic errors discussed in this section are summarized in
Fig.~\ref{fig:error_budgets}.
We see that the largest source of systematic uncertainty by far comes from the
chiral-continuum extrapolation, which includes higher-order discretization
effects. This is especially obvious at small $q^2$, i.e., large $r_1 E_K$,
because the statistical errors of the correlations functions increase with
increasing recoil momentum so that the corresponding form factors at large $r_1
E_K$ have large errors. This is also due to a lack of data points in the large
$r_1 E_K$ region as shown in Fig.~\ref{fig:fpar_fperp_NNLO}. Furthermore, the
HMrS$\chi$PT used to perform the extrapolation is valid only for moderate
$E_K$. This is a generic feature common to all similar lattice calculations.
Our aim, however, is to get the form factor in the whole kinematically allowed
region, all the way to $q^2 = 0$. In the next section, Sec.~\ref{sec:contin},
we will describe how the extrapolation can be done by including physical
information to control the error in the small $q^2$ region.

The sub-dominant errors, excluding the chiral-continuum extrapolation error,
have mild $q^2$ dependence. Following Ref.~\cite{Lattice:2015tia} we therefore
treat them as constants in $q^2$ when propagating them to the
$z$-parametrization fit in Sec.~\ref{subsec:continuum_form_factors}. We
conservatively take the maximum estimated error from each source in the
simulated $q^2$ range and add them in quadrature. Specifically, the overall
additional systematic error is 1.4\% for both $f_+$ and $f_0$, which is added
to the covariance function of the chiral-continuum fit using the procedure
described in Ref.~\cite{Lattice:2015tia} prior to the next step in the analysis
described in the following section.

%% file: section6-continuum-form-factors.tex
\section{Continuum form factors} \label{sec:contin}

The continuum form factors obtained from the chiral-continuum extrapolations
described in the previous two sections are reliable only in the high momentum
transfer $q^2 \gtrsim 17~\text{GeV}^2$ region. In this section, we use a
model-independent parametrization and expansion, namely the
$z$-parametrization, to extrapolate the form factors to the whole kinematically
allowed region. This parametrization and expansion is based on the analyticity
of the form factors and angular momentum conservation. The parametrization we
used was introduced by Bourrely, Caprini, and Lellouch
(BCL)~\cite{Bourrely:2008za} and the fitting procedure and extrapolation
technique was first introduced in our previous $\bpi$
paper~\cite{Lattice:2015tia}. 

In Sec.~\ref{subsec:z-parametrization}, we briefly review the
$z$-parametrization and give the expansion form used in the analysis. In
Sec.~\ref{subsec:continuum_form_factors}, we present the extrapolated continuum
form factors in the whole kinematically allowed region. The results are shown
in Table~\ref{tab:z_central}, and Figs.~\ref{fig:fpf0_versus_z_qsq}
and~\ref{fig:fpf0_versus_qsq_compare_lattice_only}. A comparison with results
of other groups is presented in Sec.~\ref{subsec:compare_form_factors}.

\input{./sec6/subsection-6.1-functional-z}

\input{./sec6/subsection-6.2-continuum-z}

\input{./sec6/subsection-6.3-compare-z}

%% file: sec6/subsection-6.1-functional-z.tex
\subsection{\texorpdfstring{$z$}{z} parametrization of form factors}
\label{subsec:z-parametrization}

Before discussing the details of the method, let us first consider the
properties of the semileptonic form factors. Causality and
unitarity~\cite{eden1966analytic} imply that the $\bsk$ semileptonic form
factors are real analytic functions%
\footnote{
An analytic function $f(x)$ is real analytic if it satisfies $f(x^*)
=(f(x))^*$. If $f(x)$ is a real analytic function with a branch point at $x_0$,
then $f(x)$ is real for $x < x_0$ and its discontinuity across the cut is
purely imaginary: $f(x + i\epsilon) - f(x - i\epsilon) = 2i\mathrm{Im}f(x
+i\epsilon)$.
}
in the complex $q^2$-plane with a cut from $q^2 > \tcut$ to $\infty$, except at
physical poles below $\tcut$. The parameter $\tcut$ is the
particle-pair-production threshold. For $\bsk$, this is
\beq
\sqrt{\tcut} = M_{B^+} + M_{\pi^0} = 5.414~\mathrm{GeV} .
\label{eq:tcut}
\enq
The pole for the vector form factor is below the cut; while the one for the
scalar form factor is above it. The above-threshold pole corresponds to an
unstable particle, or resonance, and may appear only on the second Riemann
sheet.


From deep-inelastic-scattering experiments and perturbative QCD
scaling~\cite{Lepage:1980fj,Akhoury:1993uw}, it is known that the semileptonic
form factors vanish rapidly as $1/q^2$, up to logarithmic corrections, when
$q^2$ approaches minus infinity.

Near the threshold $\tcut$, the form factors have the following scaling
behavior
\begin{subequations}
\begin{align}
\mathrm{Im} f_l(q^2) &\sim (q^2 - \tcut)^\frac{2l+1}{2},
\label{eq:scaling_im}
\\
\mathrm{Re} f_l(q^2) &\sim a_l + b_l(q^2 - \tcut) , 
\label{eq:scaling_re}
\end{align}
\label{eq:scaling}
\end{subequations}
with $l = 0$ for $f_0$ and $l = 1$ for $f_+$ , obtained from simple partial
wave analysis.

\begin{figure}[htp]
\centering
\includegraphics[width=0.45\columnwidth]{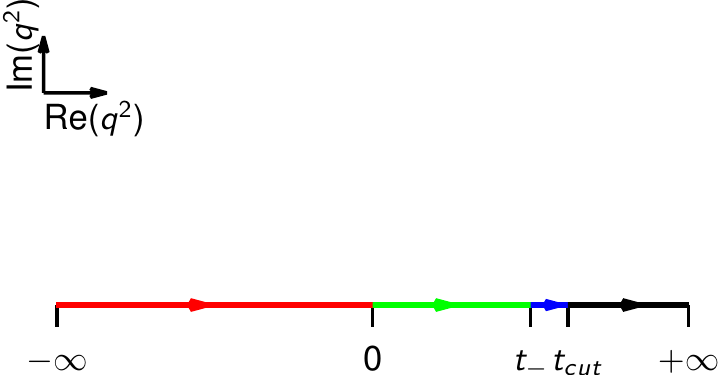}
\qquad
\includegraphics[width=0.45\columnwidth]{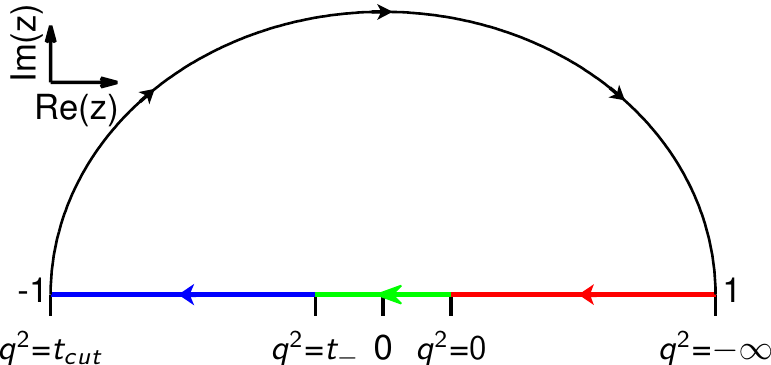}
\caption{A schematic diagram of the conformal mapping of the form factor
regions from the complex $q^2$-plane to the complex $z$-plane.}
\label{fig:conformal}
\end{figure}

Now let us look at the $z$ parametrization. The $z$ parametrization involves a
conformal mapping. Conventionally, the variable $q^2$ is mapped to a new
variable $z$ according to
\beq
z(q^2,t_0)
=
\frac{\sqrt{\tcut - q^2} - \sqrt{\tcut - t_0}}
{\sqrt{\tcut - q^2} + \sqrt{\tcut - t_0}},
\label{eq:conformal}
\enq
where $t_0$ is a parameter that can be chosen to optimize the mapping. The
maximum momentum transfer allowed in the semileptonic $\bsk$ decay is defined
as
\beq
t_- = (\MBs - M_K)^2
\label{eq:tminus}
\enq
for convenience. This conformal mapping was first considered in
Ref.~\cite{Meiman:1963} and further developed and used to get model-independent
constraints, usually called ``unitarity bounds'', on form factors in
Ref.~\cite{Okubo:1971jf}. A stronger constraint based on heavy-quark power
counting was derived in Ref.~\cite{Becher:2005bg}. The conformal transformation
Eq.~(\ref{eq:conformal}) maps the physical semileptonic region $ 0 \leq q^2
\leq t_-$ onto a small region on the real $z$ axis, the upper edge of the cut
onto the upper edge of the unit circle, the lower edge of the cut onto the
lower edge of the unit circle, the limiting points $q^2 = \pm \infty$ to $z =
1$, and $q^2 = \tcut$ to $z = -1$. The complex $q^2$ cut plane is mapped onto
the unit disk in the $z$ plane with the cut mapping onto the unit circle. The
parameter $t_0$ can be chosen such that the semileptonic region is centered
around $z$ = 0 after the conformal mapping. This is obtained by solving the
equation
\beq
z(q^2=0, t_0) = -z(q^2=t_-, t_0).
\enq
The solution for $t_0$ is 
\beq
t_0 = \tcut - \sqrt{\tcut(\tcut - t_-)} .
\label{eq:t0}
\enq
This mapping is schematically shown in Fig.~\ref{fig:conformal} with small
lepton masses ignored and with the optimized $t_0$ as defined in
Eq.~(\ref{eq:t0}).

Under the above transformation, the form factors are always in the region where
$|z| < 1$, and therefore they can be parametrized as a power series in $z$.
Since the physical semileptonic region in terms of $z$ is usually small, $|z|
\leq 0.205$ for $\bsk$, this parametrization converges quickly.
Table~\ref{tab:ranges} has a list of quantities in terms of $r_1 E_K$, $q^2$,
and $z$ parameters.
\begin{table}[tbp]
\centering
\caption{Quantities in terms of different parameters.}
\label{tab:ranges}
\begin{tabular}{crrr} 
\hline
\hline 
&$r_1E_K$   &$q^2~\mathrm{(GeV^2)}$  & \multicolumn{1}{c}{$z$}
\tabularnewline 
\hline
Lattice data range  & $[0.846, 1.71]$ & $[17.4, 23.3]$  & $[-0.186, -0.0174]$
\tabularnewline 
Physical range      & $[0.780, 4.28]$  & $[0, 23.7]$     & $[- 0.205, 0.205]$
\tabularnewline 
$t_{\_} = (\MBs - {M_K})^2$  &0.780 & 23.7            & $- 0.205$
\tabularnewline 
$\tcut = (M_{B} + {M_\pi})^2$  & $- 0.0395$    & 29.3  & $- 1.0$
\tabularnewline 
$t_0 = \tcut - \sqrt{\tcut(\tcut - t_-)}  $ &1.84 &16.5 & 0.0
\tabularnewline 
${M_{B^*}^2(1^-)}$  & 0.102           & 28.4            & $- 0.569$
\tabularnewline 
${M_{B^*}^2(0^+)}$  & $- 0.473$       & 32.3            & $- 0.625 + 0.781 i$
\tabularnewline 
\hline  
\hline  
\end{tabular} 
\end{table}

Two commonly used parametrizations are given by Boyd, Grinstein and Lebed
(BGL)~\cite{Boyd:1994tt} and by Bourrely, Caprini and Lellouch
(BCL)~\cite{Bourrely:2008za}. Here we use the BCL parametrization as given by
\begin{subequations}
\begin{align}
f_+(q^2)
&=
\frac{1}  {1-q^2/ m_{B^*(1^-)}^2}
\sum\limits_{k=0}^{K-1} b_k^+(t_0)
\left[z^k -(-1)^{k-K}\frac{ k}{ K} z^{K}\right], 
\label{eq:BCL_f+} \\
f_0(q^2)
&=
\frac{1}{1-q^2/ m_{B^*(0^+)}^2}
\sum\limits_{k= 0}^{K-1} b_k^0(t_0)\, z^k .
\label{eq:BCL_f0}
\end{align}
\label{eq:BCL}
\end{subequations}
The factors $1/({1-q^2/ m_{B^*}^2})$ take the poles into account and ensure the
asymptotic scaling, $f(q^2) \sim 1/q^2$ at large $q^2$. Moreover, the scaling
condition of Eq.~(\ref{eq:scaling}) near $\tcut$ is also enforced for $f_+$.
Note that Eq.~(\ref{eq:scaling}) in the $q^2$-plane imply the following
relation
\begin{subequations}
\begin{align}
\frac{df_+}{dz}|_{z=-1}
&=
\frac{df_+}{dk}\frac{dk}{dz}|_{k=0} = 0,
\label{eq:dfpdz}
\\
\frac{df_0}{dz}|_{z=-1}
&=
\frac{df_0}{dk}\frac{dk}{dz}|_{k=0} = \mathrm{const} .
\label{eq:df0dz}
\end{align}
\end{subequations}

The form factors constructed with this BCL $z$ parametrization satisfy all
three properties of the semileptonic form factors discussed at the beginning of
this section.

%% file: sec6/subsection-6.2-continuum-z.tex
\subsection{\texorpdfstring{$z$}{z}-parametrization fit and extrapolation}
\label{subsec:continuum_form_factors}

We use Eq.~(\ref{eq:BCL}) to perform the $z$-parametrization fit to our
chiral-continuum-extrapolated form factor results obtained in
Secs.~\ref{subsec:chiral_extrap} and \ref{sec:errors}. The vector pole
$M_{B^*(1^-)}$ is taken to be $M_{B^*(1^-)} = 5.32465(25)
\mathrm{GeV}$~\cite{Tanabashi:2018oca}, and the above threshold scalar pole
$M_{B^*(0^+)}$ is taken to be the theoretically predicted value $M_{B^*(0^+)} =
5.68 \mathrm{GeV}$~\cite{Gregory:2010gm}. The parameter $t_0$ is chosen as in
Eq.~(\ref{eq:t0}), and the corresponding value for the $\bsk$ process is
$16.5~\mathrm{GeV}^2$. Table~\ref{tab:masses} lists the relevant meson masses
used in the $z$-parametrization fit.
\begin{table}[tbp]
\centering
\caption{Input meson masses used in the $z$-parametrization fit.}
\label{tab:masses}
\begin{tabular}{ccccccc} 
\hline
\hline 
&$\MBs$   & $M_K$	& $M_B$	& $M_{\pi}$	& ${M_{B^*}(1^-)}$	& ${M_{B^*}(0^+)}$
\tabularnewline 
\hline  
Value $\mathrm{(GeV)}$ & 5.36682	& 0.493677	& 5.27931	& 0.1349766 & 5.32465	& 5.68
\tabularnewline 
\hline  
\hline  
\end{tabular} 
\end{table}

The functional method introduced in Ref.~\cite{Lattice:2015tia} is used to
perform the $z$-parametrization fit, where, following
Ref.~\cite{Lattice:2015tia}, we take as inputs the results from the
chiral-continuum extrapolation and systematic error analysis as presented in
Sec.~\ref{subsec:chiral_summary}).

Our preferred (central) fit has $K = 4$, where $K$ is the number of terms in
the expansion in Eq.~(\ref{eq:BCL}). The results of this fit are shown in
Table~\ref{tab:z_central}. These can be used to reconstruct the final form
factors as described in Appendix~\ref{app:reconstruct}. We arrive at this
preferred fit choice by first simultaneously fitting the form factors $f_+$ and
$f_0$ with $K = 2$ and without constraining the $z$-parametrization parameters
$b_i^{+, 0}$ in Eq.~(\ref{eq:BCL}). The coefficients $b_0^+$ and $b_0^0$ are
well determined, but the quality of this fit is poor. When increasing $K$ from
2 to 3, the quality of the fit improves, and all the $b_i^{+, 0}$ coefficients
can be determined well. The kinematic constraint Eq.~(\ref{eq:kin_constraint})
is satisfied within errors%
\footnote{Note that the kinematic constraint is automatically satisfied in
Eq.~(\ref{eq:fp_f0_fpar_fperp}) before taking the extrapolation as is being
done in this section. After the extrapolation, this constraint is not
guaranteed if not imposed in the fit.}.
Enforcing this kinematic constraint, as explained below, further improve the
$f_+$ form-factor fit. The fit parameters also satisfy the unitarity
condition~\cite{Bourrely:2008za} and the condition estimated from heavy-quark
power counting~\cite{Becher:2005bg}. Adding the heavy-quark constraint does not
affect the fit results. The kinematic constraint is enforced by requiring $f_+$
and $f_0$ to be exactly equal at the $q^2 = 0$ point. In practice, we set a
prior in the $z$-parametrization fit
\beq
f_+(q^2 = 0) - f_0(q^2 = 0) = 0 ,
\label{eq:kin_constraint_practice}
\enq
with width $\epsilon = 10^{-10}$. When further increasing the expansion order
to $K = 4$, the central value of the form factors at $q^2 = 0$ agrees with the
results with $K = 3$, but the error increases. The unitarity and heavy-quark
constraints are still satisfied automatically. The results stabilize at $K = 4$
and do not change with $K = 5$. We conclude that the $K = 4$ fit with the
kinematic constraint includes the systematic uncertainty due to truncating the
$z$-parametrization series. 

\begin{figure}
\centering
\includegraphics[width=0.45\columnwidth]{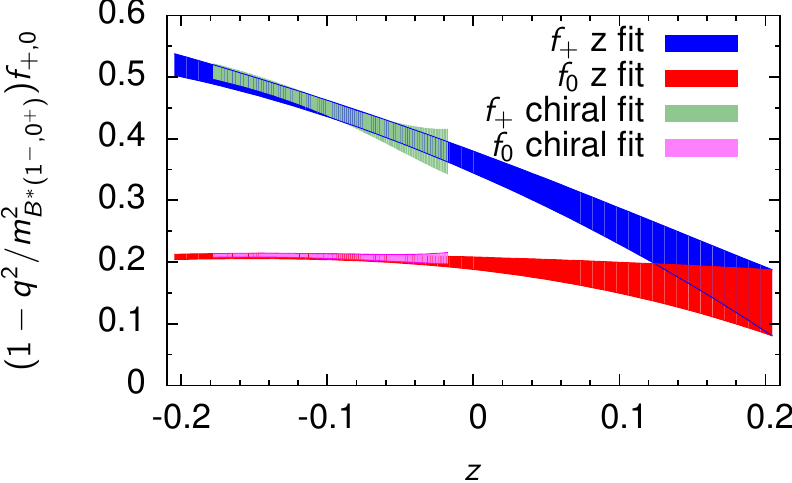}
\qquad
\includegraphics[width=0.45\columnwidth]{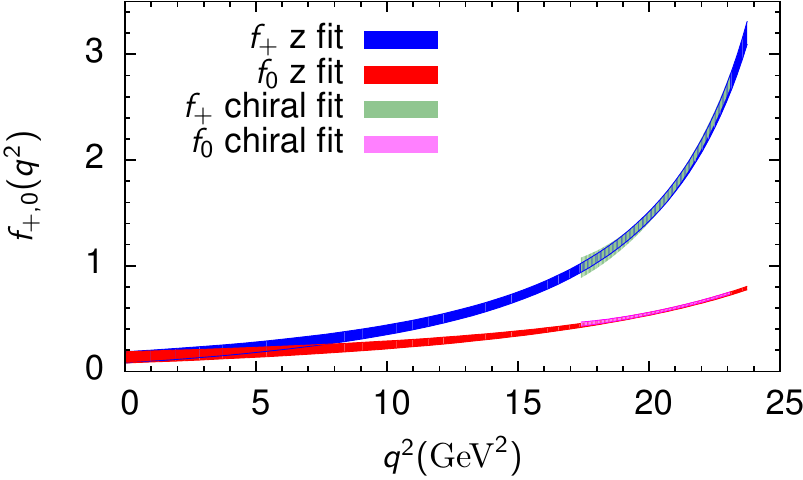}
\caption{Preferred $K = 4$ $z$-parametrization fit results for the form factors
$f_+$ (upper curve) and $f_0$ (lower curve) as functions of $z$ and $q^2$. The
kinematic constraint Eq.~(\ref{eq:kin_constraint_practice}) is applied. The
corresponding bands with larger errors are the results of the chiral-continuum
extrapolation, as shown in Sec.~\ref{subsec:chiral_extrap}. They are used as
inputs for the $z$-parametrization fit.  The bands with smaller errors are the
resultant $z$-parametrization fits. The $q^2 = 0$ point corresponds to $z =
0.205$ as shown in Table~\ref{tab:ranges}. The meson poles are listed in
Table~\ref{tab:masses}.}
\label{fig:fpf0_versus_z_qsq}
\end{figure}

The left panel of Fig.~\ref{fig:fpf0_versus_z_qsq} shows the preferred $K=4$
form-factor results, with poles removed, as functions of $z$. The $q^2 = 0$
point is at the right end of the plot.  Note that the shape of the form factors
as functions of $z$ is parametrization dependent. For convenience, the right
panel of Fig.~\ref{fig:fpf0_versus_z_qsq} shows the form factors as functions
of $q^2$. The $q^2$ dependence of the form factors is parametrization
independent and can be used directly to compare with results of other groups.
\begin{table}[tbp]
\centering
\caption{The results of the preferred $z$-parametrization fit from
Eqs.~(\ref{eq:BCL},~\ref{eq:t0},~\ref{eq:kin_constraint_practice}) and
Table~\ref{tab:masses} with $K = 4$. These values can be used to reconstruct
the form factors as explained in Appendix~\ref{app:reconstruct}. The
correlation matrix is listed with only four digits after the decimal point.
The correlation matrix has one near zero eigenvalue due to the kinematic
constraint used. See Appendix~\ref{app:reconstruct} for details.}
\label{tab:z_central}
\begin{tabular}{rcrrrrrrrr}
\hline
\hline
\multicolumn{10}{c}{Correlation matrix} \\
\cline{3-10}
&   Value       & $b_0^+$ & $b_1^+$ & $b_2^+$ & $b_3^+$ & $b_0^0$ &
$b_1^0$ & $b_2^0$ & $b_3^0$\\
\hline

 $b_0^+$ & 0.3623(0.0178)  &  1.0000 & 0.6023 & 0.0326 & -0.1288 & 0.7122 & 0.6035 & 0.5659 & 0.5516 \\
 $b_1^+$ & -0.9559(0.1307) &         & 1.0000 & 0.4735 &  0.2677 & 0.7518 & 0.9086 & 0.9009 & 0.8903 \\
 $b_2^+$ & -0.8525(0.4783) &         &        & 1.0000 &  0.9187 & 0.5833 & 0.7367 & 0.7340 & 0.7005 \\
 $b_3^+$ & 0.2785(0.6892)  &         &        &        &  1.0000 & 0.4355 & 0.5553 & 0.5633 & 0.5461 \\
 $b_0^0$ & 0.1981(0.0101)  &         &        &        &         & 1.0000 & 0.8667 & 0.7742 & 0.7337 \\
 $b_1^0$ & -0.1661(0.1130) &         &        &        &         &        & 1.0000 & 0.9687 & 0.9359 \\
 $b_2^0$ & -0.6430(0.4385) &         &        &        &         &        &        & 1.0000 & 0.9899 \\
 $b_3^0$ & -0.3754(0.4535) &         &        &        &         &        &        &        & 1.0000 \\
\hline
\hline
\end{tabular}
\end{table}

%% file: sec6/subsection-6.3-compare-z.tex
\subsection{Comparison with existing results}
\label{subsec:compare_form_factors}

\begin{figure}
\centering
\includegraphics[width=0.8\columnwidth]{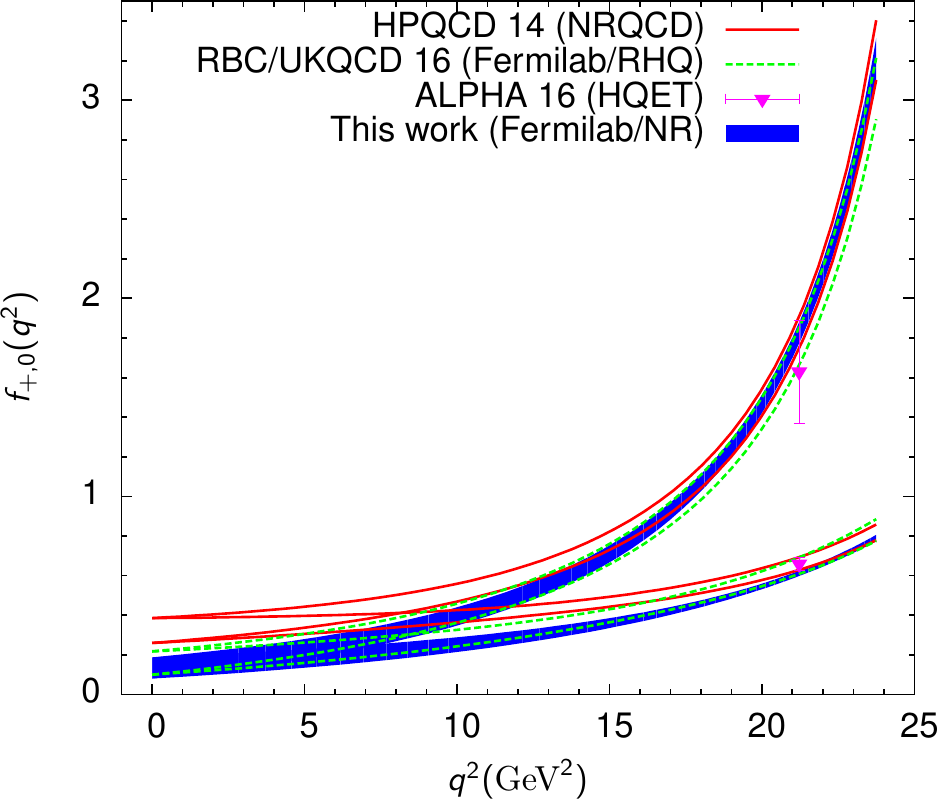}
\caption{Theoretical lattice QCD calculations of the $\bsk$ form factors from the HPQCD
Collaboration~\cite{Bouchard:2014ypa}, the RBC and UKQCD
Collaborations~\cite{Flynn:2015mha}, the ALPHA
Collaboration~\cite{Bahr:2016ayy}, and the Fermilab Lattice and MILC
Collaborations, marked as ``This work'' in the figure. Different treatments of
the bottom quark on the lattice are listed in parenthesis.} 
\label{fig:fpf0_versus_qsq_compare_lattice_only}
\end{figure}

Several other groups have also calculated the same form factors. We note that
Refs.~\cite{Bouchard:2014ypa} and~\cite{Flynn:2015mha} use the $B_s K$
threshold instead of $B\pi$ in their implementation of the $z$ parametrization.
Since the $z$-parameter, by definition (see Eq.~(\ref{eq:conformal})), depends
on the threshold ($t_{\rm cut}$), we cannot directly compare the $z$-dependence
of our form factors with those of Refs.~\cite{Bouchard:2014ypa}
and~\cite{Flynn:2015mha}. We therefore compare our form factors with those
from other lattice QCD calculations only as functions of $q^2$. This is shown in
Figure~\ref{fig:fpf0_versus_qsq_compare_lattice_only}.  

The results of the HPQCD Collaboration~\cite{Bouchard:2014ypa} are based on
(2+1)-flavor-MILC-asqtad configurations for the sea quarks, and employ the HISQ
action for the light valence quarks, and lattice NRQCD for the heavy $b$-quark.
The RBC and UKQCD Collaborations~\cite{Flynn:2015mha} use
(2+1)-flavor-domain-wall fermions for the sea quarks and light valence quarks,
and a variant~\cite{Lin:2006ur,Christ:2006us} of the Fermilab action for the
heavy $b$-quark. The ALPHA Collaboration~\cite{Bahr:2016ayy} uses leading-order
lattice HQET to get the form factors at one point, $q^2 =
22.12~\mathrm{GeV}^2$. While our results are consistent with those from
Refs.~\cite{Flynn:2015mha} and~\cite{Bahr:2016ayy}, they are in tension with
HPQCD's results~\cite{Bouchard:2014ypa}. We note that
Ref.~\cite{Bouchard:2014ypa} employs the so-called modified $z$-expansion,
where the chiral-continuum extrapolation is combined with the $z$-expansion
into one fit function by modifying the $z$-coefficients with lattice-spacing
and light-quark-mass dependent terms. This procedure may affect the shape of
the form factors. Indeed, in their calculation of the form factors for the $B
\to K \ell^+ \ell^-$ decay in Ref.~\cite{Bouchard:2013pna}, the HPQCD
Collaboration compared the form factors obtained after the modified
$z$-expansion with the results from a two-step method that is very similar to
ours, performing first a chiral-continuum extrapolation, and then a
$z$-expansion fit. While they find only small differences between the two sets
of form factors, those obtained from their implementation of the two-step
method are in better agreement with the results of Ref.~\cite{Bailey:2015dka}.
However, unlike the case at hand, the form factors of
Ref.~\cite{Bailey:2015dka} are not in significant tension with HPQCD's results
of Ref.~\cite{Bouchard:2013pna}. We see that the tension between our $\bsk$
form factor results and those of Ref.~\cite{Bouchard:2014ypa} increases with
decreasing $q^2$ to roughly 2.3$\sigma$ at $q^2=0$. The RBC and UKQCD
Collaborations~\cite{Flynn:2015mha}, on the other hand, adopt the same
procedure as we do, namely a chiral-continuum extrapolation at high $q^2$,
followed by a $z$-expansion extrapolation to $q^2=0$.

\begin{figure}
\centering
\includegraphics[width=0.45\columnwidth]{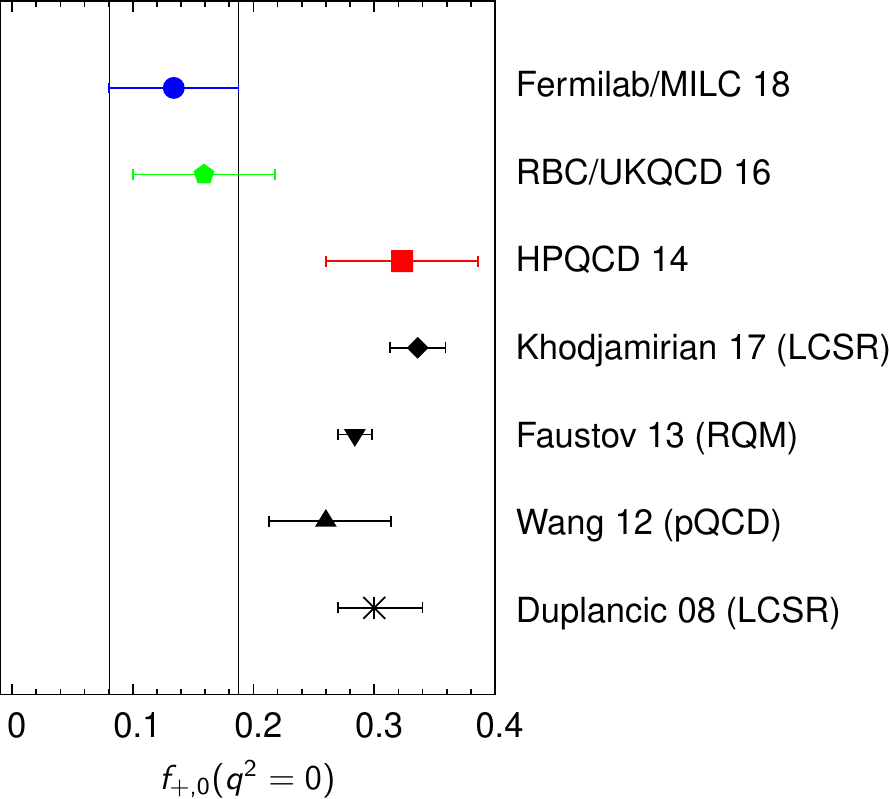}
\caption{Comparison of the theoretical calculations of the $\bsk$ form factors at $q^2 =
0$. The results shown are from light-cone sum rules
(LCSR)~\cite{Duplancic:2008tk,Khodjamirian:2017fxg}, NLO perturbative QCD
(pQCD)~\cite{Wang:2012ab}, relativistic quark model
(RQM)~\cite{Faustov:2013ima}, and (2+1)-flavor lattice QCD (LQCD) from the
HPQCD Collaboration~\cite{Bouchard:2014ypa}, the RBC and UKQCD
Collaborations~\cite{Flynn:2015mha}, and the Fermilab Lattice and MILC
Collaborations.} 
\label{fig:fpf0_versus_qsq_0_only}
\end{figure}

A comparison of the form factor at $q^2 = 0$ is shown in
Fig.~\ref{fig:fpf0_versus_qsq_0_only}, where we also include results from
calculations using light-cone sum
rules~\cite{Duplancic:2008tk,Khodjamirian:2017fxg}, a relativistic quark
model~\cite{Faustov:2013ima}, and NLO perturbative QCD~\cite{Wang:2012ab}.

%% file: section7-pheno.tex
\section{Phenomenological applications} \label{sec:pheno}

The angular-dependent differential decay rate for $\bsk$ is given in
Eq.~(\ref{eq:ddr}). One can construct at most three independent observables
from there. In the following, we will consider the differential decay rate
$d\Gamma / dq^2$ in Sec.~\ref{subsec:rate}, the forward-backward asymmetry
$A_{FB}^\ell (q^2)$ in Sec.~\ref{subsec:fb}, and the lepton polarization
asymmetry $A_{pol}^\ell(q^2)$ in Sec.~\ref{subsec:pol}. The latter two
quantities are sensitive to the mass of the final-state charged lepton. In
Sec.~\ref{subsec:ratio}, we also construct the ratios of the scalar and vector
form factors between the $\bsk$ and $\bsds$ decays.

\input{./sec7/subsection-7.1-decay-rate}

\input{./sec7/subsection-7.2-fb-asymmetry}

\input{./sec7/subsection-7.3-pol-asymmetry}

\input{./sec7/subsection-7.4-ratio}

%% file: sec7/subsection-7.1-decay-rate.tex
\subsection{Decay rate} \label{subsec:rate}

\begin{figure}[bp]
\centering
\includegraphics[width=0.45\columnwidth]{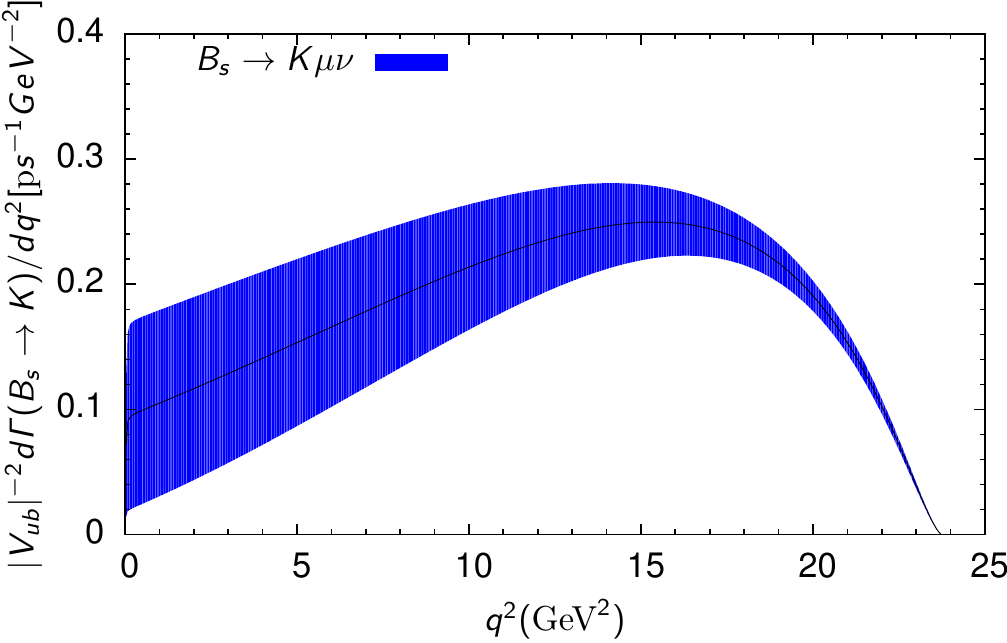}
\qquad
\includegraphics[width=0.45\columnwidth]{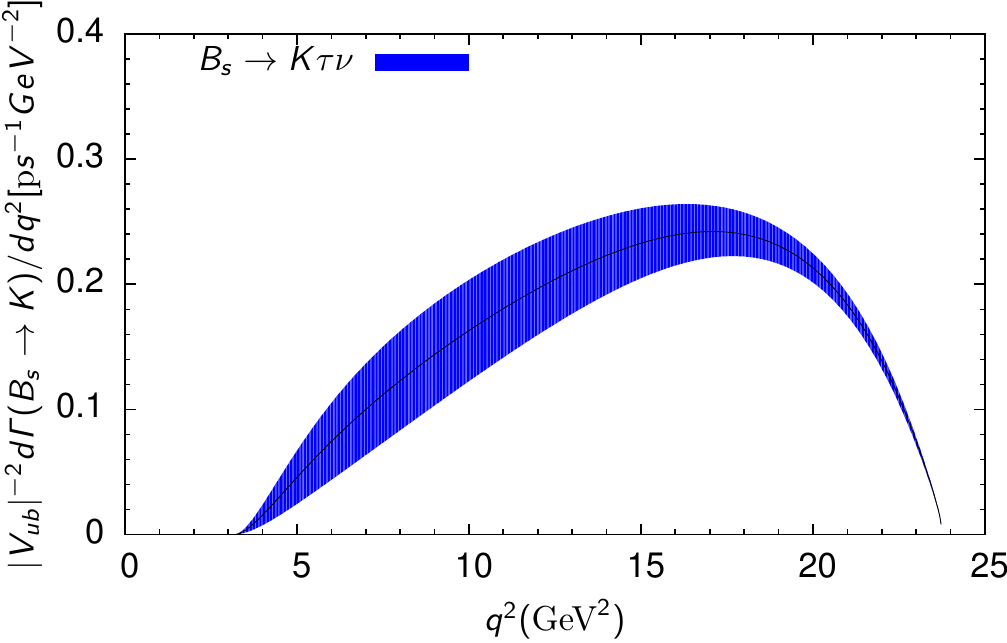}
\caption{Standard Model predictions of the differential decay rate divided by
$|V_{ub}|^2$ for $B_s \to K\mu\nu$ (left) and $B_s \to K\tau\nu$ (right).}
\label{fig:dGdqsq}
\end{figure}

The differential decay rate can be obtained from Eq.~(\ref{eq:ddr}) by
integrating over the angle $\theta_\ell$, which yields
\begin{align}
\label{eq:dGdqsq}
\frac{d\Gamma}{dq^2}
=&
\int_{-1}^{1} \frac{d^2\Gamma}{dq^2 d \cos\theta_\ell } d \cos\theta_\ell 
\nonumber\\
=&
\frac{G_F^2 |V_{ub}|^2}{128\pi^3 M_{B_s}^2}
\left( 1 - \frac{m_\ell^2}{q^2} \right)^2
|\bm{p}_K|
\left[
\frac{16}{3} M_{B_s}^2 |\bm{p}_K|^2
\left(
1 + \frac{m_\ell^2}{2q^2}  \right) 
|f_+(q^2)|^2 \right.
\\
&\left. +
\frac{2m_\ell^2}{q^2} (M_{B_s}^2 - M_K^2)^2 |f_0(q^2)|^2
\right] \nonumber .
\end{align}
In Fig.~\ref{fig:dGdqsq}, we plot the Standard Model predictions of the
differential decay rate divided by $|V_{ub}|^2$ over the whole kinematic range
of $q^2$ for $B_s \to K\mu\nu$ and $B_s \to K\tau\nu$.

One can also explore the ratio of the differential decay rates
\beq
R^{\tau/\mu}(q^2) = 
\frac{
d\Gamma(B_s \to K \tau \nu) /dq^2
}
{
d\Gamma(B_s \to K \mu \nu) /dq^2
} .
\enq
\begin{figure}[tbp]
\centering
\includegraphics[width=0.45\columnwidth]{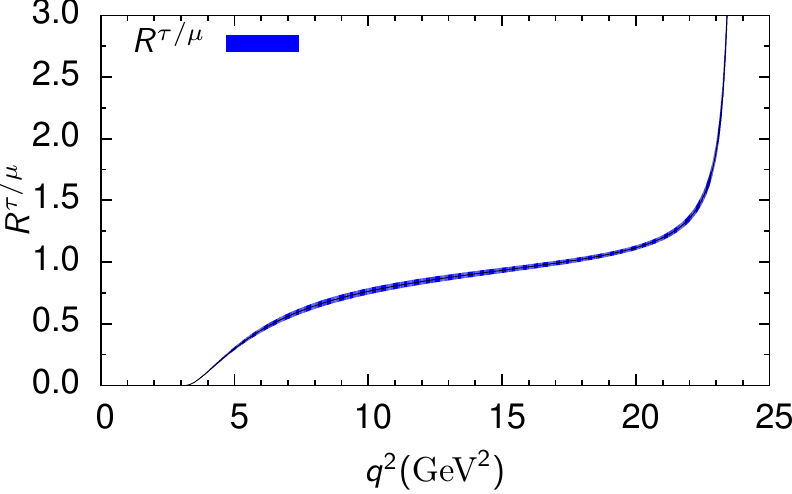}
\caption{Standard Model predictions of the ratio of the differential decay
rates $R^{\tau/\mu}(q^2)$.}
\label{fig:R_tau_mu}
\end{figure} 
Figure~\ref{fig:R_tau_mu} shows the prediction for $R^{\tau/\mu}(q^2)$.

The total decay rate is given by
\beq
\Gamma( \bsk )
=
\int_{m_\ell^2}^{q_\mathrm{max}^2}dq^2 \frac{d\Gamma}{dq^2} ,
\enq
with $q_\mathrm{max}^2 = t_- = (\MBs - M_K)^2$, as in Eq.~(\ref{eq:tminus}).
The numerical results for $\Gamma/|V_{ub}|^2$ are
\begin{subequations}
\begin{align}
|V_{ub}|^{-2} \Gamma(B_s \to K \mu  \nu) &= 4.26(0.92)~\rm{ps}^{-1}, \\
|V_{ub}|^{-2} \Gamma(B_s \to K \tau \nu) &= 3.27(0.47)~\rm{ps}^{-1} .
\end{align}
\label{eq:Gamma}
\end{subequations}

In Appendix~\ref{app:bin}, we also provide partially integrated differential
decay rates in evenly spaced $q^2$ bins. 

The ratio of the total decay rate is 
\beq
\frac{
\Gamma(B_s \to K \tau \nu)
}
{
\Gamma(B_s \to K \mu \nu)
}
= 0.836(34),
\enq
which takes the correlations between the form factors into account and is more
precise than directly using Eq.~(\ref{eq:Gamma}).

%% file: sec7/subsection-7.2-fb-asymmetry.tex
\subsection{Forward-backward asymmetry} \label{subsec:fb}

\begin{figure}[bp]
\centering
\includegraphics[width=0.45\columnwidth]{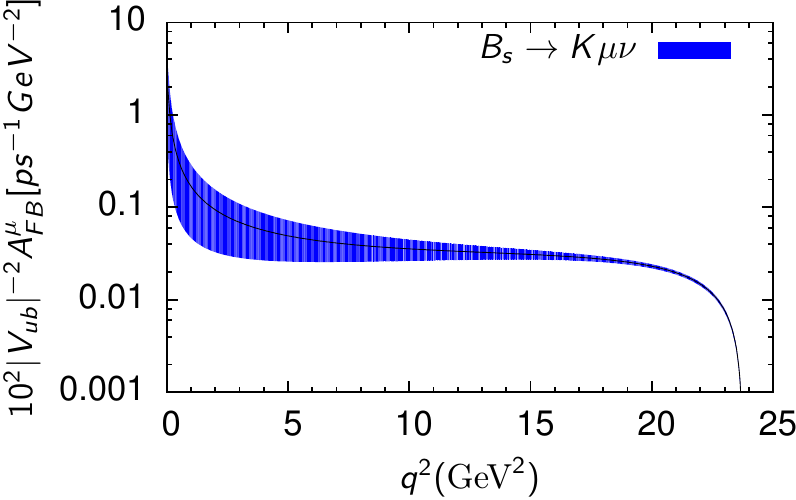}
\qquad
\includegraphics[width=0.45\columnwidth]{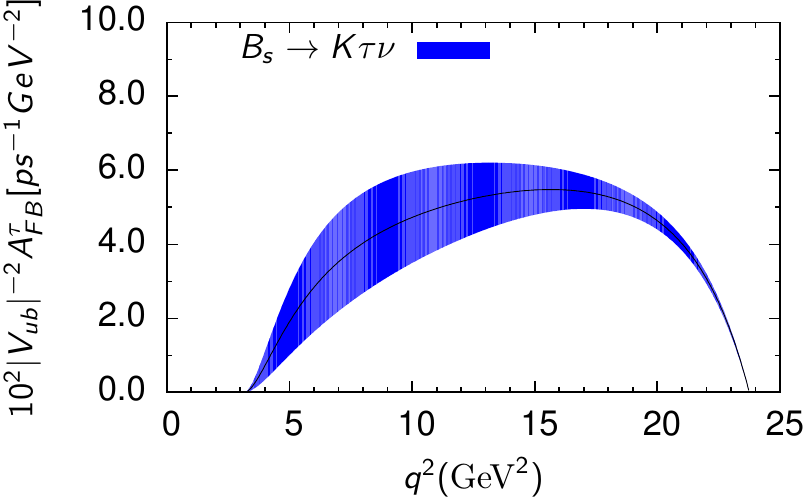}
\caption{Standard Model predictions of the forward-backward asymmetry divided by
$|V_{ub}|^2$ for $B_s \to K\mu\nu$ (left) and $B_s \to K\tau\nu$ (right).}
\label{fig:AFB_mu_NoVub}
\end{figure}

The forward-backward asymmetry, $A_{FB}$, which depends on the linear
$\cos\theta_\ell$ term in Eq.~(\ref{eq:ddr}), is given by
\begin{align}
A_{FB}^\ell (q^2)
&=
\int_0^1 \frac{d^2\Gamma}{dq^2d\cos\theta_\ell} d\cos\theta_\ell 
-
\int_{-1}^0 \frac{d^2\Gamma}{dq^2d\cos\theta_\ell} d\cos\theta_\ell 
\nonumber \\
&=
\frac{G_F^2 |V_{ub}|^2}{32\pi^3 M_{B_s}}
\left( 1 - \frac{m_\ell^2}{q^2} \right)^2
|\bm{p}_K|^2
\frac{m_\ell^2}{q^2} (M_{B_s}^2 - M_K^2)
\Re\left[f_+(q^2)  f_0^*(q^2)\right] .
\label{eq:AFB}
\end{align}
The Standard Model predictions for the forward-backward asymmetry divided by
$|V_{ub}|^2$ are shown in Fig.~\ref{fig:AFB_mu_NoVub}.
For the corresponding integrated quantities we find
\begin{subequations}
\begin{align}
\int_{m_\mu^2}^{q_\mathrm{max}^2} dq^2 |V_{ub}|^{-2} A_{FB}^\mu(q^2)
&= 0.0137(69)~\rm{ps}^{-1}, 
\\
\int_{m_\tau^2}^{q_\mathrm{max}^2} dq^2 |V_{ub}|^{-2} A_{FB}^\tau(q^2)
&= 0.83(14)~\rm{ps}^{-1} .
\end{align}
\end{subequations}

The normalized forward-backward asymmetry is given by
\beq
\bar{A}_{FB}^\ell
\equiv
\frac{
\int_{m_\ell^2}^{q_\mathrm{max}^2}
A_{FB}^\ell (q^2)
}
{
\int_{m_\ell^2}^{q_\mathrm{max}^2}
d\Gamma/dq^2
} 
\enq
and the corresponding numerical values are
\begin{subequations}
\begin{align}
\bar{A}_{FB}^\mu
&= 0.00321(97) ,
\\
\bar{A}_{FB}^\tau
&= 0.2536(84) .
\end{align}
\end{subequations}

%% file: sec7/subsection-7.3-pol-asymmetry.tex
\subsection{Lepton polarization asymmetry} \label{subsec:pol}

\begin{figure}[tbp]
\centering
\includegraphics[width=0.45\columnwidth]{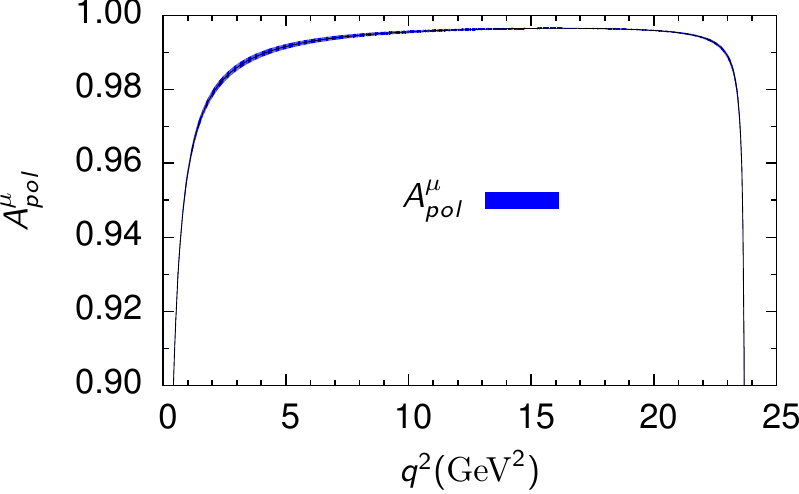}
\qquad
\includegraphics[width=0.45\columnwidth]{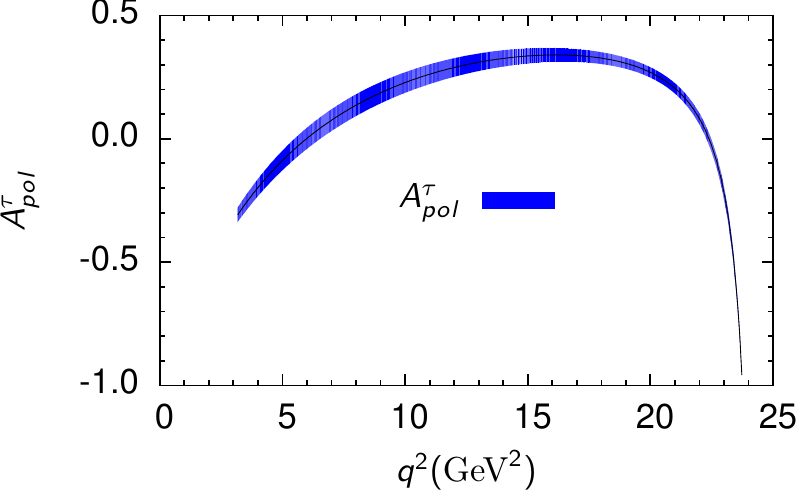}
\caption{Standard Model predictions of the normalized lepton polarization
asymmetry for $B_s \to K\mu\nu$ (left) and $B_s \to K\tau\nu$ (right).}
\label{fig:Apol}
\end{figure}

The normalized lepton polarization asymmetry is defined as
\beq
A_\text{pol}^\ell = 
\frac{
d\Gamma^-/dq^2
-
d\Gamma^+/dq^2
}
{
d\Gamma^-/dq^2
+
d\Gamma^+/dq^2
}
\enq
from the differential decay rates with definite lepton helicity
\cite{Meissner:2013pba}
\begin{subequations}
\begin{align}
\frac{d\Gamma^-}{dq^2}
&=
\frac{G_F^2 |V_{ub}|^2}{24\pi^3}
\left( 1 - \frac{m_\ell^2}{q^2} \right)^2
|\bm{p}_K|^3
|f_+(q^2)|^2 ,
 \\
\frac{d\Gamma^+}{dq^2}
&=
\frac{G_F^2 |V_{ub}|^2}{24\pi^3}
\left( 1 - \frac{m_\ell^2}{q^2} \right)^2
\frac{m_\ell^2}{q^2}
|\bm{p}_K|
\left[
\frac{3}{8}
\frac{(M_{B_s}^2 - M_K^2)^2}{M_{B_s}^2} |f_0(q^2)|^2
+
\frac{1}{2} |\bm{p}_K|^2 |f_+(q^2)|^2
\right].
\end{align}
\end{subequations}
Here the superscripts $+$ ($-$) imply a right- (left-)handed lepton in the
final state. The lepton is produced via the $V-A$ current in the Standard
Model, and therefore the electron and muon are mainly left-handed polarized.
The $A_\text{pol}^\mu$ is close to one in the whole $q^2$ range. Here we
provide the normalized lepton polarization asymmetry $A_\text{pol}^\mu$ and
$A_\text{pol}^\tau$ as functions of $q^2$ in Fig.~\ref{fig:Apol}.

%% file: sec7/subsection-7.4-ratio.tex
\subsection{Ratio of the \texorpdfstring{$\bsk$}{Bs to K l nu} and
\texorpdfstring{$\bsds$}{Bs to Ds l nu} form factors} \label{subsec:ratio}

We also calculate the ratios of the scalar and vector form factors between the
$\bsk$ and $\bsds$ semileptonic decays. The ratios can be used along with
future experimental results to determine the ratio of the CKM matrix elements
$|V_{ub}/V_{cb}|$.

\begin{figure}[tbp]
\centering
\includegraphics[width=0.45\columnwidth]{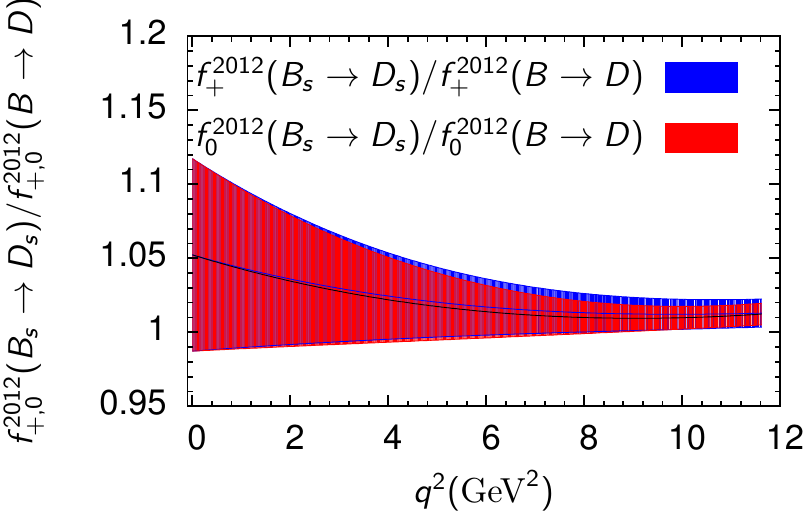}
\qquad
\includegraphics[width=0.45\columnwidth]{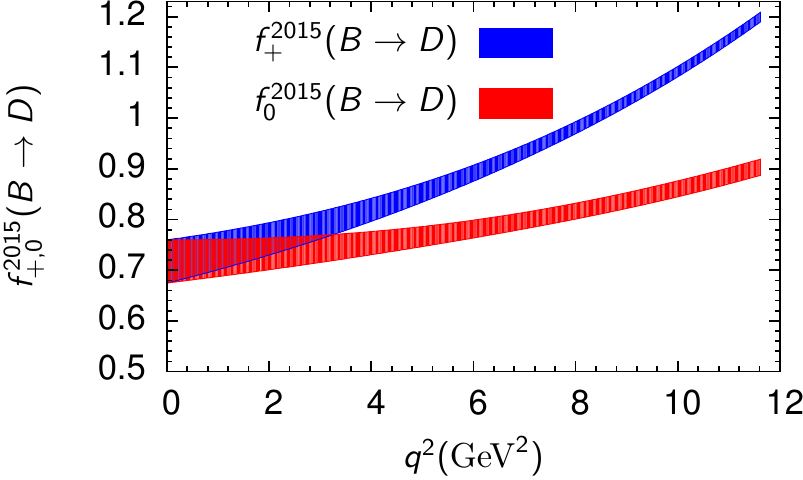}
\caption{Form factor ratios, $f_{+,0}^{2012}(B_s\to D_s) / f_{+,0}^{2012}(B\to
D)$, calculated by Fermilab Lattice and MILC Collaborations
in~\cite{Bailey:2012rr} in 2012 (left) and $\bd$ form factors,
$f_{+,0}^{2015}(B\to D)$, calculated by the same collaborations
in~\cite{Lattice:2015rga} in 2015 (right). These are the ingredients to
reconstruct the $\bsds$ form factors $f_{+,0}^{\mathrm{reco}}(B_s\to D_s)$.}
\label{fig:bs2ds_2012_b2d_2015}
\end{figure}

\begin{figure}[tbp]
\centering
\includegraphics[width=0.45\columnwidth]{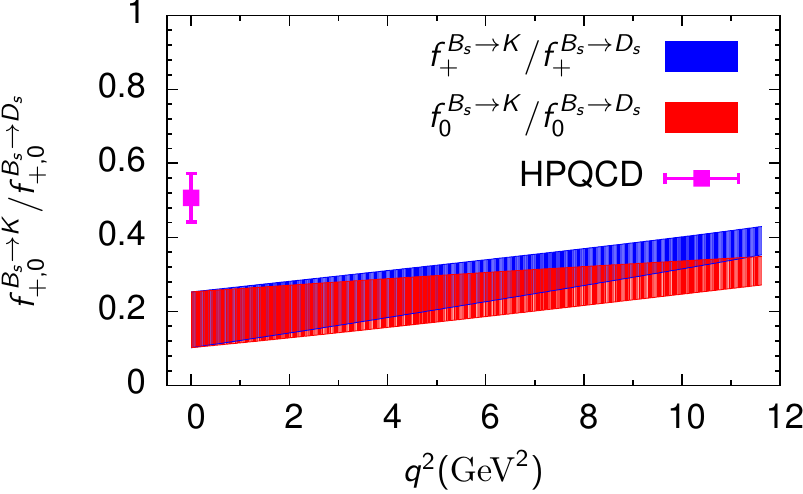}
\caption{Form factor ratios, $f_{+,0}(B_s\to K) /
f_{+,0}^{\mathrm{reco}}(B_s\to D_s)$, as functions of the momentum transfer
$q^2$. The result provided by HPQCD~\cite{Monahan:2018lzv} at $q^2 = 0$ is
plotted for comparison.}
\label{fig:bsds_bsk_ratio}
\end{figure}

\begin{figure}[tbp]
\centering
\includegraphics[width=0.45\columnwidth]{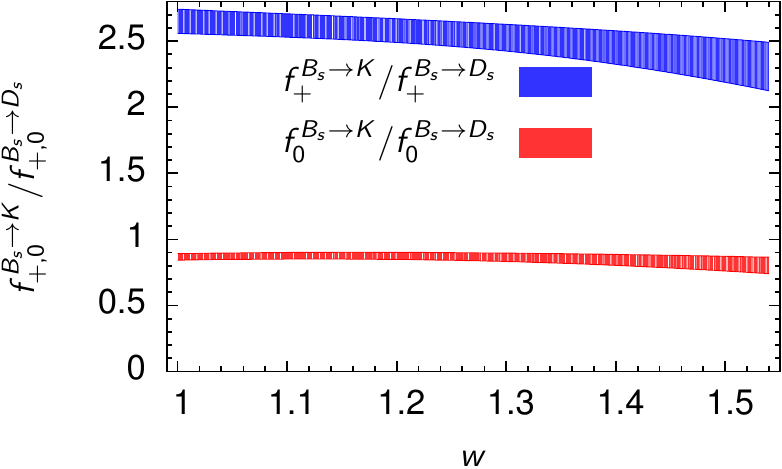}
\caption{Form factor ratios, $f_{+,0}(B_s\to K) /
f_{+,0}^{\mathrm{reco}}(B_s\to D_s)$, as functions of the recoil parameter
$w$.}
\label{fig:bsds_bsk_ratio_w}
\end{figure}

\begin{figure}[tbp]
\centering
\includegraphics[width=0.45\columnwidth]{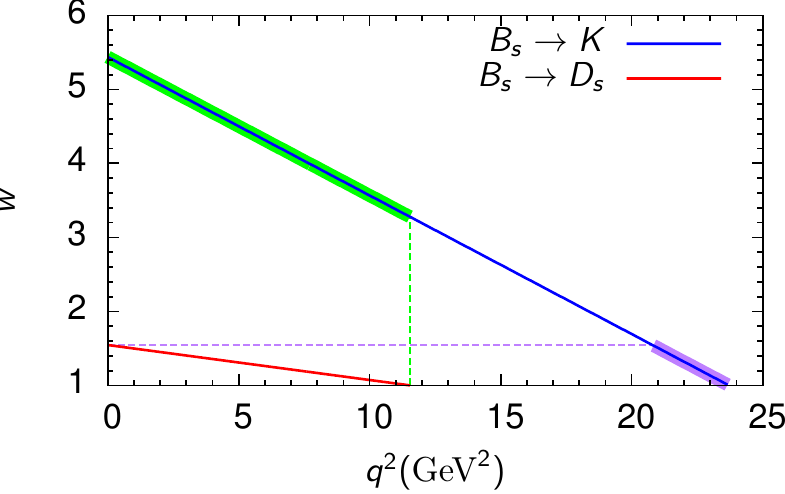}
\caption{The kinematically allowed region for $\bsk$ (upper solid line) and
$\bsds$ (lower solid line) decays in terms of $q^2$ and $w$. The solid lines
are the relation between $q^2$ and $w$ as defined in Eq.~(\ref{eq:w_qsq}). The
green and purple areas are the corresponding $\bsk$ regions used to construct
the form factor ratios as shown in Fig.~\ref{fig:bsds_bsk_ratio} and
Fig.~\ref{fig:bsds_bsk_ratio_w}, respectively.}
\label{fig:w_vs_qsq}
\end{figure}

First, we reconstruct the $\bsds$ form factors from our previous
papers~\cite{Bailey:2012rr,Lattice:2015rga}. Form factor ratios,
$f_{+,0}^{2012}(B_s\to D_s)/f_{+,0}^{2012}(B\to D)$, and the $\bd$ form
factors, $f_{+,0}^{2015}(B\to D)$, are calculated in Refs.~\cite{Bailey:2012rr}
and~\cite{Lattice:2015rga}, respectively. They are shown in
Fig.~\ref{fig:bs2ds_2012_b2d_2015}.  
\begin{figure}[tbp]
\centering
\includegraphics[width=0.45\columnwidth]{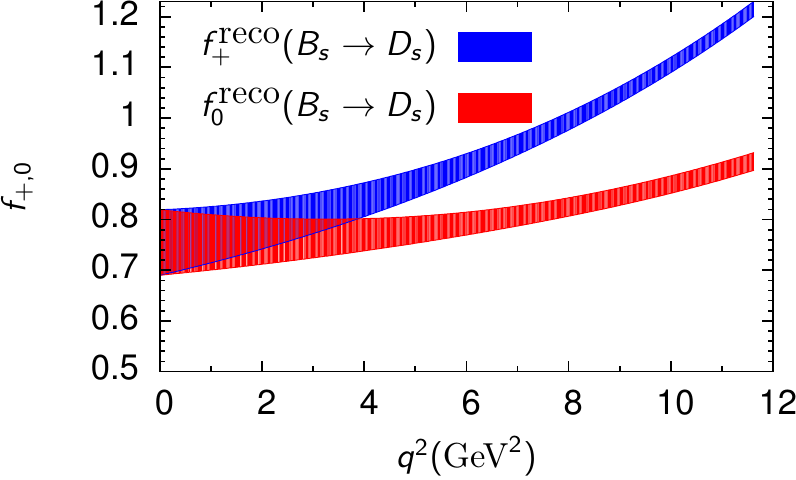}
\caption{The reconstructed form factors $f_{+,0}^{\mathrm{reco}}(B_s\to D_s)$
obtained from Eq.~(\ref{eq:reconstructed}).}
\label{fig:bs2ds_reconstructed}
\end{figure} 
The $\bsds$ form factor can be reconstructed via
\beq
f_{+,0}^{\mathrm{reco}}(B_s\to D_s) = f_{+,0}^{2015}(B\to D) \times
\frac{f_{+,0}^{2012}(B_s\to D_s)}{f_{+,0}^{2012}(B\to D)} .
\label{eq:reconstructed}
\enq
With the reconstructed $\bsds$ form factors $f_{+,0}^{\mathrm{reco}}(B_s\to
D_s)$ shown in Fig.~\ref{fig:bs2ds_reconstructed}, we obtain the form-factor
ratios, $f_{+,0}(B_s\to K) / f_{+,0}^{\mathrm{reco}}(B_s\to D_s)$, shown in
Fig.~\ref{fig:bsds_bsk_ratio} as functions of $q^2$ and in
Fig.~\ref{fig:bsds_bsk_ratio_w} as functions of $w$. Although the 2012 analysis
was carried out on a subset of the ensembles used in the 2015 analysis, we
neglect any correlations in the two form factors in
Eq.~(\ref{eq:reconstructed}). Here $q^2$ is the usual square of the lepton
momentum transfer as defined in Eq.~(\ref{eq:qsq_E}). The recoil parameter $w$
for $\bsds$ is defined as
\beq
w = \frac{M_{B_s}^2 + M_{D_s}^2 - q^2}{2M_{B_s} M_{D_s}}
\label{eq:w_qsq}
\enq
and the corresponding one for the $\bsk$ is defined by replacing $M_{D_s}$ with
$M_K$. The relation between $w$ and $q^2$  in Eq.~(\ref{eq:w_qsq}), and the
kinematically allowed regions for the two types of processes are shown in
Fig.~\ref{fig:w_vs_qsq}. The ratios constructed with different parameters $q^2$
and $w$ as shown in Figs.~\ref{fig:bsds_bsk_ratio}
and~\ref{fig:bsds_bsk_ratio_w} allow us to probe the different $\bsk$ form
factor regions.

%% file: section8-summary.tex
\section{Summary and Outlook} \label{sec:outlook}

Using six strategically selected ensembles of MILC asqtad 2+1 flavor gauge
configurations, we have calculated the form factors $f_+(q^2)$ and $f_0(q^2)$
needed to understand the semileptonic decay $\bsk$. We present predictions of
the differential decay rate (divided by $|V_{ub}|^2$) for both light ($e$ or
$\mu$) or heavy ($\tau$) final-state leptons. Once the experimental data become
available, our form factors can be used to determine $|V_{ub}|$, which can then
be compared to and, if consistent, combined with the $|V_{ub}|$ determinations
from other exclusive decay processes. Hence they may help shed light on the
discrepancy with $|V_{ub}|$ from inclusive decays $B\to X_u\ell\nu$ and,
perhaps, contribute to evidence for new physics beyond the Standard Model by
enabling more stringent tests of the CKM paradigm. Other quantities of
phenomenological interest include the forward-backward asymmetry
$A_\text{FB}^\ell(q^2)$ and the lepton polarization asymmetry
$A_\text{pol}^\ell(q^2)$. We also present ratios of the form factors $f_+$ and
$f_0$ for $\bsk$ and $\bsds$ as functions of both $q^2$ and $w$. These may be
valuable for determining $|V_{ub}/V_{cb}|$.

Although there are no published results for the decay $\bsk$, this process is
under investigation by the LHCb experiment, and will be studied by the Belle~II
Collaboration when they run at the $\Upsilon(5\text{S})$ resonance, which is a
copious source of $B_s$ and $\bar B_s$ mesons.

On the theoretical side, we have plans to reduce the contributions from the
dominant sources of systematic errors in upcoming calculations, which include
chiral extrapolation, light and heavy-quark discretization, and
renormalization. The gauge ensembles generated by the MILC collaboration with
four flavors of HISQ sea quarks~\cite{Bazavov:2010ru,Bazavov:2012xda} are a
crucial ingredient in these plans. These ensembles cover a lattice spacing
range of approximately 0.15--0.045 fm with physical light quark masses and a
dynamical charm quark. The chiral extrapolation becomes a chiral interpolation,
and the reduced taste breaking of the HISQ action greatly reduces light quark
discretization errors. Using these ensembles, we will be taking two approaches
to the $b$ quark. First, we have started a project using Fermilab $b$ quarks
(as in this project) and HISQ light valence quarks.  Preliminary results were
already reported in Refs.~\cite{Gelzer:2017edb} and~\cite{Lattice2018}. As a
further small improvement compared to this work, we will include the full
correlation matrix between form factors for different processes in our final
results. In our second approach, we plan to use the HISQ formalism for the $b$
quark to calculate semileptonic $B_{(s)}$- and $D$-meson decay form factors
again on the HISQ ensembles. Heavy-quark discretization errors are simpler with
the HISQ action than with the Fermilab approach, and can be controlled with
high precision by including ensembles with very fine lattice spacings in the
range of $a\approx 0.03-0.042$~fm. The heavy-HISQ approach also allows us to
take advantage of Ward identities when renormalizing the currents. Indeed, our
recent work~\cite{Bazavov:2017lyh} employing the heavy HISQ method for the $B$-
and $D$-meson decay constants has reached unprecedented precision. We have
recently started to generate the correlation functions for this project. In
summary, with the improvements outlined above, we expect, in the coming years,
to obtain the form factors for $\bsk$ (and related decays) with percent level
precision, at least in the low recoil region of the phase space. 

%% file: acknowledgements.tex
\begin{acknowledgments}

We thank Jon A. Bailey for participating in the early stage of the project. We
thank Chris M. Bouchard for discussions and comments on the form factor
comparison section. We thank members of the LHCb Collaboration for discussions
and in particular Svende Braun, Marta Calvi, and Mika A. Vesterinen for sharing
with us their analysis status and the preferred $q^2$ bins. 

Computations for this work were carried out with resources provided by
the USQCD Collaboration,
the National Energy Research Scientific Computing Center,
the Argonne Leadership Computing Facility, 
the Blue Waters sustained-petascale computing project,
the National Institute for Computational Science,
the National Center for Atmospheric Research,
the Texas Advanced Computing Center,
and Big Red II+ at Indiana University.
USQCD resources are acquired and operated thanks to funding from the Office of Science of the U.S. Department of Energy.
%
The National Energy Research Scientific Computing Center is a DOE Office of Science User Facility supported by the
Office of Science of the U.S. Department of Energy under Contract No.\ DE-AC02-05CH11231.
%
An award of computer time was provided by the Innovative and Novel Computational Impact on Theory and Experiment (INCITE)
program. This research used resources of the Argonne Leadership Computing Facility, which is a DOE Office of Science
User Facility supported under Contract DE-AC02-06CH11357.
%
The Blue Waters sustained-petascale computing project is supported by the National Science Foundation (awards OCI-0725070 and
ACI-1238993) and the State of Illinois.
Blue Waters is a joint effort of the University of Illinois at Urbana-Champaign and its National Center for Supercomputing
Applications.
This work is also part of the ``Lattice QCD on Blue Waters'' and ``High Energy Physics on Blue Waters'' PRAC allocations supported
by the National Science Foundation (award numbers 0832315 and 1615006).
This work used the Extreme Science and Engineering Discovery Environment (XSEDE), which is supported by National Science Foundation
grant number ACI-1548562~\cite{XSEDE_REF}.
Allocations under the Teragrid and XSEDE programs included resources at the National Institute for Computational Sciences (NICS) at
the Oak Ridge National Laboratory Computer Center, The Texas Advanced Computing Center and the National Center for Atmospheric
Research, all under NSF teragrid allocation TG-MCA93S002.
Computer time at the National Center for Atmospheric Research 
was provided by NSF MRI Grant CNS-0421498, NSF MRI Grant CNS-0420873, NSF MRI Grant CNS-0420985, NSF sponsorship of the National
Center for Atmospheric Research, the University of Colorado, and a grant from the IBM Shared University Research (SUR) program.
Computing at Indiana University is supported by Lilly Endowment, Inc., through its support for the Indiana University Pervasive
Technology Institute.

This project was supported in part by the URA Visitig Scholar Award 12-S-15
(Y.L.); 
by the U.S.\ Department of Energy under grants
No.~DE-FG02-91ER40628 (C.B.),
No.~DE-FC02-12ER41879 (C.D.),
No.~DE-FG02-13ER42001 (A.X.K.),
No.~DE{-}SC0015655 (A.X.K., Z.G.), 
No.~DE{-}SC0010120 (S.G.),     
No.~DE{-}FG02-91ER40661 (S.G.),
No.~DE-SC0010113 (Y.M.), 
No.~DE-SC0010005 (E.T.N.),
No.~DE-FG02-13ER41976 (D.T.); 
by the U.S.\ National Science Foundation under grants
PHY14-14614 and PHY17-19626 (C.D.),
and PHY14-17805~(J.L.);
by the MINECO (Spain) under grants FPA2013-47836-C-1-P and FPA2016-78220-C3-3-P (E.G.);
by the Junta de Andaluc\'{\i}a (Spain) under grant No.\ FQM-101 (E.G.);
by the Fermilab Distinguished Scholars program (A.X.K.); 
by the German Excellence Initiative and the European Union Seventh Framework Program under grant agreement No.~291763 as well as 
the European Union's Marie Curie COFUND program (A.S.K.).
Brookhaven National Laboratory is supported by the United States Department of Energy, Office of Science, Office of High Energy
Physics, under Contract No.\ DE{-}SC0012704.
This document was prepared by the Fermilab Lattice and MILC Collaborations using the resources of the Fermi National Accelerator Laboratory (Fermilab), a U.S. Department of Energy, Office of Science, HEP User Facility. Fermilab is managed by Fermi Research Alliance, LLC (FRA), acting under Contract No. DE-AC02-07CH11359.
\end{acknowledgments}

%% file: appendix.tex
\appendix

\input{./appex/appendix-su2-chiral}

\input{./appex/appendix-reconstruct}

\input{./appex/appendix-decay-rate-bin-table}

%% file: appex/appendix-su2-chiral.tex
\section{\texorpdfstring{$\bsk$}{Bs to K l nu} form factors in SU(2) chiral
perturbation theory} \label{app:su2-chiral}

In this appendix, we derive Eq.~(\ref{eq:chiral_f_nlo}), the SU(2) chiral
formula for $\bsk$.

We start from the next-to-leading order (NLO) SU(3) HMrS$\chi$PT expression for
$B_x\to P_{xy}$ semileptonic decay. It is expressed
as~\cite{Aubin:2007mc}
\begin{align}
f_{P,\mathrm{NLO}}^{B_x\to P_{xy}} 
=& 
f_{P}^{(0)}
[ 
c^{0}_{P}(1+\delta f_{P,\mathrm{logs}})+
c^{x}_{P}m_x+
c^{y}_{P}m_y+ 
c^{\mathrm{sea}}_{P}(m_u+m_d+m_s)+ \nonumber\\
&
c^{E}_{P}{E}+
c^{E^2}_{P}{E}^{2}+
c^{a^2}_{P}{a^{2}}
],
\label{eq:chiral_f_su3} 
\end{align}
where the subscript $P$ stands for $\parallel$ or $\perp$; $c^i_P$ are
coefficients and the corresponding rescaled quantities in
Eq.~(\ref{eq:chi_lhEa2}) will be determined by the chiral fits; $\delta f_{P,
\mathrm{logs}}$ contains the one-loop nonanalytic contributions and
wave-function renormalizations; $m_x$ and $m_y$ are the corresponding valence
quark masses; $m_u$, $m_d$ and $m_s$ are sea quark masses; $E = p\cdot v$ is
the $P_{xy}$ meson energy in the $B_x$ meson rest frame; and $a$ is the lattice
spacing. The leading order terms for $\fpar$ and $\fperp$ are 
\beq
f_{P}^{(0)} = \frac{1}{f}\frac{g}{E + \Delta_{xy,P}^* +  D_\mathrm{logs}},
\enq
where $f$ is the decay constant involved; $g$ is the coupling constant;
$\Delta_{xy,P}^*$ is the mass difference between the quantum number $J^P = 0^+$
or $1^-$ $B_y^*$ meson and the pseudoscalar $B_x$ meson masses at leading order
in the chiral expansion, i.e., $\Delta_{xy}^* = B_y^* - B_x$; and
$D_\text{logs}$ is the nonanalytic self-energy contribution. The scalar pole
was not included in Ref.~\cite{Aubin:2007mc} as the $0^+$ meson is not in
leading order HMrS$\chi$PT. It is added here phenomenologically as explained in
Sec.~\ref{subsec:chiral_extrap}.

For the $\bsk$ analysis considered here, $x=s^\prime$ and
$y=u^\prime=d^\prime=u=d$. Here we use primed quantities to denote the valence
quarks and the unprimed for the sea quarks\footnote{This is different from the
convention used in the main text. For example, in Tables~\ref{tab:ensembles}
and~\ref{tab:valence} the prime quantities denote the sea quarks and the
unprimed for the valence ones.}. All the data generated for $\bsk$ analysis are
partially quenched points, i.e., $m^\prime_{s} \neq m_s$. 
\begin{table}[tbp]
\centering
\caption{Fixed parameters used in the chiral-continuum extrapolation fit
function. The $r_{1}^{2}a^{2}\Delta_{\xi}$ with $\xi=P,A,T,V,I$ and
$r_{1}^{2}a^{2}\delta'_{V/A}$ are taste splittings and hairpin parameters.}
\label{tab:taste_splits}
\begin{tabular}{ccccc} 
\hline
\hline 
$\approx$a~(fm)  & 0.12  & 0.09  & 0.06  & 0\\
\hline
$r_1\mu$  & 6.831904  & 6.638563  & 6.486649  & 6.015349\\
\hline 
$r_{1}^{2}a^{2}\Delta_{P}$    & 0       & 0       & 0       & 0 \\
$r_{1}^{2}a^{2}\Delta_{A}$    & 0.22705 & 0.07469 & 0.02635 & 0\\
$r_{1}^{2}a^{2}\Delta_{T}$    & 0.36616 & 0.12378 & 0.04298 & 0\\ 
$r_{1}^{2}a^{2}\Delta_{V}$    & 0.48026 & 0.15932 & 0.05744 & 0\\ 
$r_{1}^{2}a^{2}\Delta_{I}$    & 0.60082 & 0.22065 & 0.07039 & 0\\ 
$r_{1}^{2}a^{2}\delta'_{V}$   & 0.0     & 0.0     & 0.0     & 0\\ 
$r_{1}^{2}a^{2}\delta'_{A}$   & $-0.28$   & $-0.09$   & $-0.03$   & 0\\ 
\hline
\hline  
\end{tabular} 
\end{table}

In the SU(2) limit, the $s$-quark mass is treated as infinitely heavy and all
the explicit $m_s$ dependent terms are removed from the formula. However, one
will still need to keep the mass difference, $m^\prime_{s} - m_s$, in the
leading-order analytic term to take the partial quenching effects into account.
In the hard-kaon limit, the kaon with a large energy, $E$, is integrated out in
the nonanalytic chiral expressions. These two limits greatly simplify the
expressions of the chiral logs. Following the recipes presented in the Appendix
of Ref.~\cite{Bailey:2015dka}, we obtain the SU(2) hard-kaon chiral log terms
in $\fpar$ and $\fperp$ for the $\bsk$ as
\begin{subequations}
\begin{align}
\delta f_{P, \mathrm{logs}}^{\mathrm{SU(2)}}
=&
\frac{1}{(4\pi f)^2}
\Biggl\{
\frac{1}{16}\sum_{\xi}
\left[
- I_1(m_{\pi,\xi})
\right]
+ \frac{1}{4}I_1(m_{\pi,I})
+
I_1(m_{\pi,V}) -I_1(m_{\eta,V})\nonumber\label{eq:logs_su2_deltaf}\\ 
&{}
+ [V\to A]\Biggr\}, \\
D_{P, \mathrm{logs}}^{\mathrm{SU(2)}}
&=
0.
\end{align}
\label{eq:logs_su2}
\end{subequations}
The summation $\xi$ is over 16 staggered fermion tastes (P, V, T, A, or I);
$I_1(m)$ is the chiral logarithm defined as 
\beq
I_1(m) = m^2\ln(\frac{m^2}{\Lambda^2}).
\enq 
Meson masses for the 2+1 case in the SU(2) limit ($m_u = m_d$ and $m_s
\rightarrow \infty$) are~\cite{Aubin:2003mg, Bailey:2015dka}
\begin{subequations}
\begin{align}
m^2_{\pi,\xi} &= m_{uu,\xi}^2 = m_{dd,\xi}^2,\\
m^2_{\eta,V(A)} &=  m_{uu,V(A)}^2 + \frac{1}{2}a^2\delta_{V(A)}^\prime.
\label{eq:metasq}
\end{align}  
\end{subequations}
The $[V\to A]$ in Eq.~(\ref{eq:logs_su2_deltaf}) stands for terms with
subscripts changed from V to A. The hairpin parameters $\delta_{V(A)}^\prime$
in Eq.~(\ref{eq:metasq}) are listed in Table~\ref{tab:taste_splits}. The
$m^2_{ij,\xi}$ are defined later in Eq.~(\ref{eq:tree_mass}).

We can regroup relevant terms in Eq.~(\ref{eq:chiral_f_su3}), drop the
$m_s^\prime$ dependent term due to the SU(2) limit, and write the formula as
the following
\begin{align}
f_{P, \mathrm{NLO}}
=& 
f_{P}^{(0)}
[ 
c^{0}_{P}(1+\delta f_{P,\mathrm{logs}})+
\frac{(c^{u}_{P} +2c^{\mathrm{sea}}_{P})}{3} 3m_u+
\nonumber
\\
&
c^{\text{sea}}_{P}(m_s-m_{s^\prime})+ 
c^{E}_{P}{E}+
c^{E^2}_{P}{E}^{2}+
c^{a^2}_{P}{a^{2}}
].
\end{align}

We can further write all the expansion parameters in terms of dimensionless
ones
\begin{subequations}
\begin{align}
\chi_\text{l} & = \frac{3(2\mu m_{u})}{8\pi^{2}f^{2}},\\
\chi_\text{h} & = \frac{2\mu(m_{s}-m^\prime_{s})}{8\pi^{2}f^{2}},\\
\chi_{E} & = \frac{\sqrt{2}E}{4\pi f},\\
\chi_{a^{2}} & = \frac{a^{2}\bar{\Delta}}{8\pi^{2}f^{2}},\label{eq:chi_a2}
\end{align}
\label{eq:chi_lhEa2}
\end{subequations}
where $\mu$ is the leading-order low-energy constant that relates the
tree-level mass of a taste-$\xi$ meson composed of quarks of flavor $i$ and $j$
to the corresponding quark masses
\beq
m^2_{ij,\xi} = \mu(m_i + m_j) + a^2\Delta_\xi.
\label{eq:tree_mass}
\enq
Here $\Delta_\xi$ is the staggered fermion taste splitting. The numerical
values of $\mu$ and $\Delta_\xi$ are determined by the MILC Collaboration and
are shown in Table~\ref{tab:taste_splits}. The average taste splitting in
Eq.~(\ref{eq:chi_a2}) is $\bar{\Delta} = \frac{1}{16}\sum_\xi \Delta_\xi$.

Combining the above information, one arrives at the final NLO form used in the
chiral-continuum extrapolation in this work, Eq.~(\ref{eq:chiral_f_nlo}).

%% file: appex/appendix-reconstruct.tex
\section{Reconstructing the \texorpdfstring{$\bsk$}{Bs to K l nu} form factors}
\label{app:reconstruct}

In this appendix, we document the procedure of reconstructing the form factors
from the fitting results obtained in Sec.~\ref{subsec:continuum_form_factors}.

\subsection{Reconstructing the form factors as functions of
\texorpdfstring{$z$}{z}} \label{app_subsec:z}

The form factors are parametrized in a BCL~\cite{Bourrely:2008za} form with
coefficients $b_i^{+,0}$ as shown in Eq.~(\ref{eq:BCL}). The meson masses used
in the $z$-parametrization fit are listed in Table~\ref{tab:masses}. The fitted
coefficients $b_i^{+,0}$ are listed in Table~\ref{tab:z_central}. To get the
form factors as functions of $z$ and reproduce the left panel result of
Fig.~\ref{fig:fpf0_versus_z_qsq}, one should use Eq.~(\ref{eq:BCL}) with
${M_{B^*}(1^-)}$ and ${M_{B^*}(0^+)}$ meson mass values in
Table~\ref{tab:masses}, and the $b_i^{+,0}$ values and the correlation matrix
in Table~\ref{tab:z_central}.

\subsection{Reconstructing the form factors as functions of
\texorpdfstring{$q^2$}{q-squared}}

To get the $q^2$ dependence of the form factors as in the right panel of
Fig.~\ref{fig:fpf0_versus_z_qsq}, one needs the relation between $z$ and $q^2$.
In this paper, the mapping is defined in
Eqs.~(\ref{eq:conformal}),~(\ref{eq:tcut}),~(\ref{eq:tminus}),
and~(\ref{eq:t0}). One can then solve Eq.~(\ref{eq:conformal}) to get $q^2$ in
terms of $z$:
\beq
q^2(z,t_0)
=
\tcut - \left(
\frac{1 + z}
{1 - z}
\right)^2
\left(
\tcut - t_0
\right) .
\label{eq:conformal_z2qsq}
\enq
Once we have the form factors as functions of $z$ from
Appendix~\ref{app_subsec:z}, we can then use Eq.~(\ref{eq:conformal_z2qsq}) to
change the variable to get the $q^2$ dependence.

\subsection{Dealing with the near zero eigenvalue in the covariance matrix}

In Table~\ref{tab:z_central} the fit parameter standard deviations and the
correlation matrix are listed. To get the covariance matrix, one only needs to
follow the usual procedure to rescale the correlation matrix. The following is
the detailed procedure.

Suppose the standard deviation of the fit parameters is 
\beq
\Sigma = \left[\sigma_1, \sigma_2, \cdots, \sigma_n \right] , 
\enq
and the matrix $D$ is a diagonal matrix with diagonal elements $\Sigma$. The
correlation matrix is denoted as $R$ and the covariance matrix is denoted as
$S$. The relations among $D$, $R$, and $S$ are
\begin{subequations}
\begin{align}
S &= D \times R \times D , \\
R &= D^{-1} \times S \times D^{-1} .
\end{align}
\end{subequations}
Alternatively, one can use the following relation to directly convert the
matrix elements
\beq
S_{ij} = R_{ij} \sigma_i \sigma_j ,
\enq
where there is no summation over the repeated indices. The covariance matrix,
or the inverse of it, is useful when combining form factor results from
different sources. It is difficult to calculate the inverse of the covariance
matrix from the results listed in Table~\ref{tab:z_central}. This is because we
imposed the kinematical constraint Eq.~(\ref{eq:kin_constraint_practice}) with
$\epsilon = 10^{-10}$ in the $z$-parametrization fit. This results in a near
zero eigenvalue in the covariance matrix. The kinematic constraint is
equivalent to reducing one parameter in the $z$ parametrization. In principle,
one can first reduce one parameter, say $b_3^0$, in Eq.~(\ref{eq:BCL}), express
it in terms of the other $b_i^{+,0}$ parameters, and then perform the
$z$-parametrization fit. This, however, will make the expression
Eq.~(\ref{eq:BCL_f0}) cumbersomely complicated to handle when performing the
fit. In practice, we use the expressions Eq.~(\ref{eq:BCL}) and perform the
fits as described in Sec.~\ref{subsec:continuum_form_factors}. Whenever one
needs to invert the covariance matrix, one simply needs to reduce the size of
the matrix by removing one column and one row corresponding to one parameter
$b_r$. The parameter $b_r$ can be any one of the $b_i^{+,0}$ parameters.
Without loss of generality, let us pick $b_r$ to be $b_{K-1}^0 = b_3^0$ for our
$K = 4$ preferred fit. From Eqs.~(\ref{eq:BCL}
and~\ref{eq:kin_constraint_practice}), we can get
\begin{align}
b_{K-1}^0(t_0) =& \sum\limits_{k=0}^{K-2} 
\left[
\left(b_k^+(t_0) - b_k^0(t_0)\right) z^{k - K + 1}
-
(-1)^{k - K} \frac{k}{K} z b_k^+(t_0)
\right]  \\
& + b_{K-1}^+(t_0) \left(1 + \frac{K - 1}{K}z\right), \nonumber
\end{align}
with
\beq
z \equiv z(q^2 = 0,t_0)
=
\frac{\sqrt{\tcut} - \sqrt{\tcut - t_0}}
{\sqrt{\tcut} + \sqrt{\tcut - t_0}} ,
\enq
as derived from Eq.~(\ref{eq:conformal}).

%% file: appex/appendix-decay-rate-bin-table.tex
\section{\texorpdfstring{$\bsk$}{Bs to K l nu} differential decay rate bin
tables} \label{app:bin}

In this appendix, we present the quantity 
\beq
\frac{1}{|V_{ub}|^{2}} \int_{q_1^2}^{q_2^2}dq^2 \frac{d\Gamma}{dq^2} 
\label{eq:qsq_bins}
\enq
for the $B_s\to K\mu\nu$ and $B_s\to K\tau\nu$ decays, in bins of $q^2$, in
Tables~\ref{table:bsk_bf_bin_mu} and~\ref{table:bsk_bf_bin_tau}. Since we also
include the correlations between $q^2$ bins in these tables, the results
therein can be combined with the corresponding experimental measurements to
determine $|V_{ub}|$. 

\begin{sidewaystable}[tbp]
\caption{\label{table:bsk_bf_bin_mu}%
The binned differential decay rates, defined in Eq.~(\ref{eq:qsq_bins}), and
their correlations for $B_s \to K \mu \nu$ in twelve evenly spaced $q^2$ bins.
}
\centering
\begin{ruledtabular}
\begin{tabular}{cc*{22}{D{.}{.}{1.4}}}
\multicolumn{14}{c}{Correlation matrix} \\
\cline{3-14}
$\Delta q^2(\mathrm{GeV}^2)$ & $\textrm{Value}$ &\multicolumn{1}{c}{0-2}& \multicolumn{1}{c}{2-4}& \multicolumn{1}{c}{4-6}& \multicolumn{1}{c}{6-8}& \multicolumn{1}{c}{8-10}& \multicolumn{1}{c}{10-12}& \multicolumn{1}{c}{12-14}& \multicolumn{1}{c}{14-16}& \multicolumn{1}{c}{16-18}& \multicolumn{1}{c}{18-20}& \multicolumn{1}{c}{20-22}& \multicolumn{1}{c}{22-24} \\ 
\hline
 0-2   & 0.20707(0.14609) &  1.0000 & 0.9981 & 0.9911 & 0.9766 & 0.9510 & 0.9096 & 0.8459 & 0.7520 & 0.6171 & 0.4280 & 0.1869 & -0.0067 \\
 2-4   & 0.25741(0.14256) &   	    & 1.0000 & 0.9974 & 0.9880 & 0.9681 & 0.9332 & 0.8765 & 0.7893 & 0.6599 & 0.4715 & 0.2200 &  0.0060 \\
 4-6   & 0.30678(0.13329) &         &        & 1.0000 & 0.9965 & 0.9836 & 0.9564 & 0.9082 & 0.8297 & 0.7076 & 0.5217 & 0.2600 &  0.0237 \\
 6-8   & 0.35666(0.12140) &         &        &        & 1.0000 & 0.9952 & 0.9773 & 0.9396 & 0.8720 & 0.7599 & 0.5790 & 0.3085 &  0.0484 \\
 8-10  & 0.40477(0.10755) &         &        &        &        & 1.0000 & 0.9933 & 0.9683 & 0.9147 & 0.8160 & 0.6440 & 0.3674 &  0.0830 \\
 10-12 & 0.44785(0.09239) &         &  	     &        &        &        & 1.0000 & 0.9906 & 0.9545 & 0.8742 & 0.7169 & 0.4396 &  0.1323 \\
 12-14 & 0.48121(0.07653) &         &        &        &        &        &        & 1.0000 & 0.9862 & 0.9306 & 0.7967 & 0.5285 &  0.2033 \\
 14-16 & 0.49783(0.06056) &         &        &        &        &        &        &        & 1.0000 & 0.9778 & 0.8804 & 0.6382 &  0.3069 \\
 16-18 & 0.48688(0.04502) &         &        &        &        &        &        &        &        & 1.0000 & 0.9585 & 0.7714 &  0.4589 \\
 18-20 & 0.43098(0.03061) &         &        &        &        &        &        &        &        &        & 1.0000 & 0.9170 &  0.6736 \\
 20-22 & 0.30246(0.01787) &         &        &        &        &        &        &        &        &        &        & 1.0000 &  0.9090 \\
 22-24 & 0.08453(0.00502) &         &        &        &        &        &        &        &        &        &        &        &  1.0000 \\
\end{tabular}
\end{ruledtabular}
\end{sidewaystable}
\begin{sidewaystable}[tbp]
\caption{\label{table:bsk_bf_bin_tau}%
The binned differential decay rates, defined in Eq.~(\ref{eq:qsq_bins}), and
their correlations for $B_s \to K \tau \nu$ in eleven evenly spaced $q^2$ bins. 
}
\centering
\begin{ruledtabular}
\begin{tabular}{cc*{20}{D{.}{.}{1.4}}}
\multicolumn{13}{c}{Correlation matrix} \\
\cline{3-13}
$\Delta q^2(\mathrm{GeV}^2)$ & $\textrm{Value}$ & \multicolumn{1}{c}{2-4}& \multicolumn{1}{c}{4-6}& \multicolumn{1}{c}{6-8}& \multicolumn{1}{c}{8-10}& \multicolumn{1}{c}{10-12}& \multicolumn{1}{c}{12-14}& \multicolumn{1}{c}{14-16}& \multicolumn{1}{c}{16-18}& \multicolumn{1}{c}{18-20}& \multicolumn{1}{c}{20-22}& \multicolumn{1}{c}{22-24} \\ 
\hline
 2-4   & 0.00500(0.00269)   &   1.0000 &   0.9993 &   0.9953 &   0.9851 &   0.9648 &   0.9281 &   0.8640 &   0.7522 &   0.5581 &   0.2683 &   0.0393 \\
 4-6   & 0.09085(0.04127)   &          &   1.0000 &   0.9982 &   0.9909 &   0.9740 &   0.9414 &   0.8819 &   0.7747 &   0.5837 &   0.2913 &   0.0544 \\
 6-8   & 0.19913(0.07215)   &          &          &   1.0000 &   0.9971 &   0.9857 &   0.9595 &   0.9074 &   0.8081 &   0.6229 &   0.3277 &   0.0797 \\
 8-10  & 0.28718(0.08104)   &          &          &          &   1.0000 &   0.9956 &   0.9779 &   0.9359 &   0.8477 &   0.6718 &   0.3761 &   0.1156 \\
 10-12 & 0.36067(0.07799)   &          &          &          &          &   1.0000 &   0.9931 &   0.9643 &   0.8915 &   0.7302 &   0.4384 &   0.1660 \\
 12-14 & 0.42097(0.06852)   &          &          &          &          &          &   1.0000 &   0.9885 &   0.9370 &   0.7986 &   0.5192 &   0.2379 \\
 14-16 & 0.46455(0.05576)   &          &          &          &          &          &          &   1.0000 &   0.9786 &   0.8763 &   0.6252 &   0.3429 \\
 16-18 & 0.48267(0.04190)   &          &          &          &          &          &          &          &   1.0000 &   0.9554 &   0.7621 &   0.4978 \\
 18-20 & 0.45879(0.02887)   &          &          &          &          &          &          &          &          &   1.0000 &   0.9162 &   0.7121 \\
 20-22 & 0.36241(0.01781)   &          &          &          &          &          &          &          &          &          &   1.0000 &   0.9256 \\
 22-24 & 0.14148(0.00609)   &          &          &          &          &          &          &          &          &          &          &   1.0000 \\
\end{tabular}
\end{ruledtabular}
\end{sidewaystable}